\documentclass[aps,prd,twocolumn,preprintnumbers,showpacs,nofootinbib,amssymb]{revtex4}
\usepackage{graphicx}
\usepackage{amsmath}
\usepackage{amssymb}
\usepackage{amsfonts}
\usepackage{bm}
\usepackage{color}

\def\be{\begin{equation}}
\def\ee{\end{equation}}
\def\bea{\begin{eqnarray}}
\def\eea{\end{eqnarray}}

\begin{document}

\title{Relativistic self-gravitating Bose-Einstein condensates \\
and cold baryons with a stiff equation of state}
\author{Pierre-Henri Chavanis}
\affiliation{Laboratoire de Physique Th\'eorique,
Universit\'e Paul Sabatier, 118 route de Narbonne  31062 Toulouse, France}

\begin{abstract}
Because of their superfluid properties, some compact astrophysical objects such
as neutron stars may contain a significant part of their matter in the form of a
Bose-Einstein condensate (BEC). We consider a partially-relativistic model
of self-gravitating BECs where the relation between the pressure and the
rest-mass density is assumed to be quadratic (as in the case of classical BECs)
but pressure effects are taken into account in the relation between the energy
density and the rest-mass density. At high densities, we get a stiff equation of
state similar to the one considered by Zel'dovich (1961) in the context of
baryon stars in which the baryons interact through a vector meson field. We
determine the maximum mass of general relativistic BEC stars described by this
equation of state by using the formalism of Tooper (1965). This maximum mass is
slightly larger than the maximum mass obtained by Chavanis and Harko (2012)
using a fully-relativistic model. We also consider the possibility that dark
matter is made of BECs and apply the partially-relativistic model of BECs to
cosmology. In this model, we show that the  universe experiences a stiff matter
phase, followed by a dust matter phase, and finally by a dark energy phase
(equivalent to a cosmological constant). The same evolution is obtained in
Zel'dovich (1972) model which assumes that initially, near the cosmological
singularity, the universe is filled with cold baryons. Interestingly, the
Friedmann equations can be solved analytically in that case and provide a simple
generalization of the $\Lambda$CDM model.  We point out, however, the
limitations of the partially-relativistic model for BECs and show the need
for a fully-relativistic one.
\end{abstract}


\maketitle

\section{Introduction}

Bose-Einstein condensates (BEC) play a major role in condensed matter physics
\cite{bec}. Recently, it has been suggested that they could play an important
role in astrophysics and cosmology also. Indeed, dark matter halos could
be quantum objects made of BECs. The wave properties of dark matter may
stabilize the system against gravitational collapse providing halo cores instead
of cuspy profiles that are predicted by the cold dark matter (CDM) model
\cite{nfw} but not observed \cite{observations,salb}. The resulting  coherent
configuration may be understood as the ground state of some gigantic bosonic
atom where the boson particles are condensed in a single macroscopic quantum
state $\psi({\bf r})$. In the BEC  model, the formation of dark matter
structures at small scales is suppressed by quantum mechanics.  This property
could
alleviate the problems of the CDM model such as the cusp problem
\cite{observations}  and the missing satellite problem \cite{satellites}. At the
scale of galaxies, Newtonian gravity can be used so the evolution of the wave
function $\psi({\bf r},t)$ is governed by the Gross-Pitaevskii-Poisson (GPP)
system. The Gross-Pitaevskii (GP) equation \cite{gross,pitaevskii} is valid at
$T=0$ which is relevant for the most compact dwarf halos. Using the Madelung
\cite{madelung} transformation, the GP equation turns out to be equivalent to
hydrodynamic (Euler) equations involving an isotropic pressure due to
short-range
interactions (scattering) and an anisotropic quantum pressure arising
from the Heisenberg uncertainty principle. At large scales, quantum
effects are negligible and one recovers the classical hydrodynamic
equations of the CDM model which are remarkably
successful in explaining the large-scale structure of the universe \cite{ratra}.
At small-scales, the pressure arising from the Heisenberg uncertainty
principle or from the repulsive scattering of the bosons may stabilize dark
matter halos against gravitational collapse and lead to smooth core densities
instead of cuspy density profiles in agreement with the observations
\cite{observations}. Quantum mechanics may therefore be a way to solve the
problems of the CDM model.

The possibility that dark matter could be in the form of BECs has a long history
(see recent reviews in \cite{revueabril,revueshapiro,bookspringer}). In some
works
\cite{baldeschi,membrado,sin,schunckpreprint,matosguzman,guzmanmatos,hu,mu,
arbey1,silverman1,matosall,silverman,bmn,sikivie,mvm,lee09,ch1,lee,prd1,prd2,
mhh,ch2,ch3},
it is assumed that the bosons have no self-interaction. In that case,
gravitational collapse is prevented by the Heisenberg uncertainty principle
which is equivalent to a quantum pressure. This leads to a mass-radius relation
$MR=9.95\hbar^2/Gm^2$ \cite{rb,membrado,prd2}. In order to account for the mass
and size of dwarf dark matter halos, the mass of the bosons must be extremely
small, of the order of $m\sim 2.57\times 10^{-20}\, {\rm eV}/c^2$ (see Appendix
D of \cite{clm}). Ultralight scalar fields like axions may have such small
masses (multidimensional string theories predict the existence of bosonic
particles down to masses of the order of $m\sim 10^{-33}\, {\rm eV}/c^2$). This
corresponds to ``fuzzy cold dark matter'' \cite{hu}. In other works
\cite{leekoh,peebles,goodman,arbey,lesgourgues,bohmer,prd1,prd2,briscese,harko,
pires,rmbec,rindler,lora,lensing,glgr1},  it is assumed that the bosons have
a repulsive self-interaction measured by the scattering length $a_s>0$. In that
case, gravitational collapse is prevented by the pressure arising from the
scattering. In the Thomas-Fermi (TF) approximation which amounts to neglecting
the quantum pressure, the resulting structure is equivalent to a polytrope of
index $n=1$ with an equation of state $P=2\pi \hbar ^{2}a_s\rho^2/m^{3}$
\cite{revuebec}. Its radius is given by $R=\pi(a_s\hbar^2/Gm^3)^{1/2}$
\cite{prd1,goodman,arbey,bohmer}, independent on its mass $M$. In order to
account for the size of dwarf dark matter halos, the ratio between the
mass and the scattering length of the bosons is fixed at $({\rm
fm}/a_s)^{1/3}(mc^2/{\rm eV})=0.654$ (see Appendix D of \cite{clm}). For
$a_s=10^6\, {\rm fm}$, corresponding to the value of the scattering
length observed in terrestrial BEC experiments \cite{revuebec}, this gives a
boson mass $m=65.4\, {\rm eV/c^2}$  much larger than the mass $m\sim 2.57\times
10^{-20}\, {\rm eV}/c^2$ required in the absence of
self-interaction.\footnote{Actually, using the constraint $4\pi
a_s^2/m<1.25\, {\rm cm}^2/{\rm g}$ set by the Bullet Cluster \cite{bullet},
implying $(a_s/{\rm fm})^{2}({\rm eV}/m c^2)<1.77\times 10^{-8}$, one finds the
upper bounds $m=1.69\times 10^{-2}\, {\rm eV}/c^2$  (in agreement
with the limit $m<1.87\, {\rm eV}/c^2$ obtained from cosmological
considerations \cite{limjap}) and $a_s=1.73\times 10^{-5}\, {\rm fm}$. For a
value of the boson mass $m=1.69\times 10^{-2}\, {\rm eV}/c^2$, we have $T\ll
T_c$ for all the dark matter halos so they can be considered to be at $T=0$
\cite{prep}. They are made of a solitonic core surrounded by a halo of scalar
radiation.} This may be more realistic from
a particle physics point of view. The general mass-radius relation of
self-gravitating BECs at $T=0$ with an arbitrary scattering length $a_s$,
connecting the non-interacting limit ($a_s=0$) to the TF limit ($G M^2 m
a_s/\hbar^2\gg 1$), has been determined analytically and numerically in
\cite{prd1,prd2}. These papers also provide the general density profile of dark
matter halos interpreted as self-gravitating BECs at $T=0$ (solitons). The
effect of a finite temperature has been considered in
\cite{ir,ingrosso,nikolic,msepl,madarassy,slepian,harko3,rm} using different
approaches.

Since atoms like $^7{\rm Li}$ have negative scattering lengths in terrestrial BEC experiments
\cite{revuebec}, it may be relevant to consider the possibility of
self-gravitating BECs with an attractive self-interaction ($a_s<0$). In that
case, there exist a maximum mass $M_{max}=
1.01\hbar/\sqrt{|a_s|Gm}=5.07{M_P}/\sqrt{|\lambda|}$, where $\lambda=8\pi a_s m
c/\hbar$ is the self-interaction constant and $M_P=(\hbar c/G)^{1/2}$ is the
Planck mass,  above which the BEC collapses \cite{prd1,prd2}. In most
applications, this mass is extremely small (when $|\lambda|\sim 1$ it is of the
order of the Planck mass $M_P=2.18\times 10^{-8}\, {\rm kg}$!)  so that the
collapse of the BEC is very easily realized in the presence of an attractive
self-interaction. This may lead to the formation of supermassive black holes at
the center of galaxies \cite{bookspringer}. On the other hand, when the BEC
hypothesis is applied in a cosmological context, an attractive self-interaction
can enhance the Jeans instability and accelerate the formation of structures in
the universe \cite{chavaniscosmo}.

Self-gravitating BECs have also been proposed to describe boson stars \cite{kaup,rb,thirring,breit,takasugi,colpi,bij,gleiserseul,gleiser,seidel90,kusmartsev,kusmartsevbis,leepang,jetzer,seidel94,balakrishna,schunckliddle,mielkeschunck,torres2000,wang,mielke,guzmanbh,chavharko}. For these compact objects, we must use general relativity and couple the Klein-Gordon equation to the Einstein field equations. Initially, the study of boson stars was motivated by the
axion field, a pseudo-Nambu-Goldstone boson of the Peccei-Quinn phase transition, that was proposed as a possible solution to the strong CP
problem in QCD. In the early works of Kaup \cite{kaup} and Ruffini \& Bonazzola \cite{rb}, it was assumed that the bosons have no self-interaction.  This leads to a maximum mass of boson stars equal to $M_{Kaup}=0.633M_P^2/m$. Above that mass no equilibrium configuration exists. In that case, the system collapses into a black hole. This maximum mass is much smaller than the maximum mass $M_{OV}=0.376 M_P^3/m^2$ of fermion stars determined by Oppenheimer and Volkoff \cite{ov} in general relativity. They differ by a factor
$m/M_P\ll 1$. This is because boson stars are stopped from collapsing by
Heisenberg's uncertainty principle while, for fermion stars,
gravitational collapse is avoided by Pauli's exclusion principle.  For $m\sim 1{\rm GeV}/c^2$, corresponding to the typical mass of the neutrons, the Kaup mass $M_{Kaup}\sim 10^{-19}M_{\odot}$ is very small.  This corresponds to mini boson stars like axion black holes. The mass of these mini boson stars may be too small to be astrophysically relevant. They could play a role,
however, if they exist in the universe in abundance or if the axion
mass is extraordinary small  leading to
macroscopic objects with a mass $M_{Kaup}$ comparable to the mass of
the sun (or even larger) \cite{mielke}. For example, axionic boson
stars could account for the mass of MACHOs (between $0.3$ and $0.8$
$M_{\odot}$) if the axions have a mass $m\sim 10^{-10}{\rm eV}/c^2$
\cite{mielkeschunck}. It has also been proposed that stable boson stars with a boson mass $m\sim 10^{-17}{\rm eV}/c^2$ could mimic supermassive black holes ($M\sim 10^6\, M_{\odot}$, $R\sim 10^7\, {\rm km}$) that reside at the center of galaxies \cite{torres2000,guzmanbh}. On the other hand, Colpi {\it et al.} \cite{colpi} assumed that the bosons have a repulsive self-interaction. In the Thomas-Fermi approximation, this leads to a maximum mass $M_{max}=0.0612\, \sqrt{\lambda} M_P^3/m^2$ which, for $\lambda\sim 1$, is of the order of the maximum mass of fermion stars $M_{OV}=0.376 M_P^3/m^2$. The self-interaction has the same effect on the bosons as the exclusion principle on the fermions. It plays the role of an interparticle repulsion (for $\lambda>0$) that  dominates over uncertainty pressure and prevents catastrophic gravitational collapse. Therefore, for  $m\sim 1{\rm GeV}/c^2$ and $\lambda\sim 1$, we get a maximum mass of the order of the solar mass $M_{\odot}$, similar to the mass of neutron stars, which is much larger than the maximum mass  $M_{Kaup}\sim 10^{-19}M_{\odot}$ obtained in the absence of self-interaction (an interpolation formula giving the maximum mass for any value of the self-interaction constant $\lambda$ is given in Appendix B.5 of \cite{prd1}). Therefore, a self-interaction can significantly change the physical dimensions of boson stars, making them much more astrophysically interesting.  For example, stellar mass boson stars could constitute a part of dark matter \cite{colpi,mielkeschunck}.

Recently, Chavanis and Harko \cite{chavharko} have proposed that, because of the
superfluid properties of the core of neutron stars, the neutrons (fermions)
could form Cooper pairs and behave as bosons of mass $2m_n$, where $m_n=0.940\,
{\rm Gev/c^2}$ is the mass of the neutrons. Therefore, neutron stars could
actually be BEC stars! Since the maximum mass of BEC stars $M_{max}=0.0612\,
\sqrt{\lambda} M_P^3/m^2=0.307\, {\hbar c^2\sqrt{a_s}}/{(Gm)^{3/2}}$ depends on
the self-interaction constant $\lambda$ (or scattering length $a_s$), this
allows
to overcome the (fixed) maximum mass of neutron stars $M_{OV}=0.376\, 
M_P^3/m^2=0.7\,  M_\odot$ determined by Oppenheimer and Volkoff \cite{ov}  by
modeling a neutron star as an ideal gas of fermions of mass $m_n$ (the
corresponding radius is $R=9.36 \, GM_{OV}/c^2=9.6\, {\rm km}$ and the
corresponding density is $\rho=5\times 10^{15}\, {\rm g/cm^3}$). By
taking a
scattering length of the order of $10-20\, {\rm fm}$ (hence $\lambda/8\pi\sim
95.2-190$), we obtain a maximum mass of the order of $2M_{\odot}$, a central
density of the order $1-3\times 10^{15}\, {\rm g/cm^3}$, and a radius in the
range $10-20\, {\rm km}$. This could account for the recently observed neutron
stars with masses in the range of $2-2.4\, M_{\odot}$
\cite{Lat,Dem,black1,black2,black3} much larger than the  Oppenheimer-Volkoff
limit \cite{ov}. For $M>M_{max}$, nothing prevents the gravitational collapse of
the star which becomes a black hole. On the other hand, for a boson mass of the
order of $m\sim 1 \, {\rm MeV/c^2}$ and a self-interaction constant 
$\lambda\sim 1$, we get $M_{max}\sim 10^{6}\, M_{\odot}$ and $R_{min}\sim 10^7
\, {\rm km}$. These parameters are reminiscent of supermassive black holes in
active galactic nuclei, so that stable self-interacting boson stars with $m\sim
1 \, {\rm MeV/c^2}$ could be an alternative to black holes at the center of
galaxies \cite{schunckliddle}.

Self-gravitating BECs may also find applications in the physics of black holes \cite{bookspringer}.  It has been proposed recently that microscopic quantum black holes could be BECs of gravitons stuck at a critical point \cite{dvali,casadio}. These results can be understood easily in terms of the Kaup mass and Kaup radius \cite{bookspringer}. Therefore, self-gravitating BECs can have many applications in astrophysics, cosmology and black hole physics with promising perspectives.

In this paper, we come back to certain approximations that have
been made in the study of self-gravitating BECs and discuss them in more detail.

In their study of general relativistic BEC stars, Chavanis and Harko
\cite{chavharko} first presented qualitative arguments giving the fundamental
scalings of the maximum mass $M_{*}\sim \hbar c^2\sqrt{a_s}/(Gm)^{3/2}$, minimum
radius $R_{*}\sim (a_s\hbar^2/Gm^3)^{1/2}$, and maximum density $\rho_{*}\sim
{m^3c^2}/{2\pi a_s\hbar^2}$ of BEC stars. Then, they developed two models in
order to obtain the numerical values of the prefactors. They first developed a
semi-relativistic model in which gravity is treated by general relativity
using the Tolman-Oppenheimer-Volkoff (TOV) equation but the relation between the
pressure and the energy density is given by the quadratic equation of state
$P=2\pi \hbar ^{2}a_s\epsilon^2/m^{3}c^4$ obtained from the classical
Gross-Pitaevskii equation after identifying the energy density with the
rest-mass density ($\epsilon=\rho c^2$). This is a particular case of a
polytropic equation of state, corresponding to an index $n=1$, studied by Tooper
\cite{tooper1} in general relativity. This semi-relativistic model leads to
a maximum mass $M_{max}=0.5001\, \hbar c^2\sqrt{a_s}/(Gm)^{3/2}$. This treatment
is approximate first because the energy density is not always dominated by the
rest-mass density and also because the relation between the pressure and the
rest-mass density is altered by relativistic effects. Chavanis and Harko
\cite{chavharko} also developed  a fully-relativistic model in which the
relation between the pressure and the energy density is obtained from the
Klein-Gordon equation \cite{colpi}. In the dense core, the equation of state
reduces to $P\sim \epsilon/3$ which is similar to the equation of state of the
radiation or to the equation of state that prevails in the core of neutron stars
modeled as an ideal gas of fermions at $T=0$. In the envelope, we recover the
equation of state $P=2\pi \hbar ^{2}a_s\epsilon^2/m^{3}c^4$ of a classical BEC.
This fully-relativistic model leads to a maximum mass $M_{max}=0.307\, \hbar
c^2\sqrt{a_s}/(Gm)^{3/2}$. This is the correct value of the maximum mass of BEC
stars. In this paper, we shall compare these results with a
partially-relativistic model of self-gravitating BECs where the relation
between the pressure and the rest-mass density is assumed to be given by 
$P=2\pi \hbar ^{2}a_s\rho^2/m^{3}$  (as for a classical BEC) but pressure
effects are taken into account in the relation between the energy density and
the rest-mass density ($\epsilon=\rho c^2+P$). This is a particular case of an
equation of state studied by Tooper \cite{tooper2} in general relativity.  In
the dense core, the equation of state reduces to $P\sim\epsilon$. This is a
stiff equation of state for which the velocity of sound
$c_s=\sqrt{P'(\epsilon)}c$ is equal to the velocity of light ($c_s=c$). This
type of equation of state was introduced by  Zel'dovich \cite{zeldovich} in the
context of baryon stars in which the baryons interact through a vector meson
field. In the envelope, we recover the equation of state $P=2\pi \hbar
^{2}a_s\epsilon^2/m^{3}c^4$ of a classical BEC.  This partially-relativistic
model leads to a maximum mass $M_{max}=0.4104\, \hbar
c^2\sqrt{a_s}/(Gm)^{3/2}$ intermediate between the two models considered by
Chavanis and Harko \cite{chavharko}. This treatment is, however, approximate
because the relation between the pressure and the rest-mass density is altered
by relativistic effects.

Self-gravitating BECs have also been considered in cosmology. Harko
\cite{harkocosmo} and Chavanis \cite{chavaniscosmo} independently developed
cosmological models in which dark matter is made of BECs. They solved the
Friedmann equations by assuming that the equation of state relating the pressure
to the energy density is given by $P=2\pi \hbar ^{2}a_s\epsilon^2/m^{3}c^4$, as
for Newtonian BECs. However, this equation of state is not valid when the BEC is
strongly relativistic. Therefore, their approach gives wrong results in the very
early
universe where relativistic effects are important. A fully-relativistic
model should use the equation of state derived from the Klein-Gordon
equation \cite{colpi}. This will be considered in a future work. As an
intermediate step, we consider here a partially-relativistic model in which
the relation between the pressure and the rest-mass density is assumed to be
given by  $P=2\pi \hbar ^{2}a_s\rho^2/m^{3}$  (as for a classical BEC) but
pressure effects are taken into account in the relation between the energy
density and the rest-mass density ($\epsilon=\rho c^2+P$).  This leads to a
cosmological model where the universe experiences a stiff matter phase, followed
by a dust matter phase, and finally by a dark energy phase (equivalent to a
cosmological constant). The same evolution is obtained in Zel'dovich
\cite{zeldocosmo} model which assumes that initially, near the cosmological
singularity, the universe is filled with cold baryons. Interestingly, the
Friedmann equations can be solved analytically in that case and  provide a
simple generalization of the $\Lambda$CDM model. We point out, however, the
limitations of this partially-relativistic model for BECs and the need for a
fully-relativistic one. Although our relativistic treatment
is approximate for BECs, it is exact for the
type of particles considered by Zel'dovich \cite{zeldovich,zeldocosmo}.

The paper is organized as follows. In Sec. \ref{sec_sgbec}, we recall the basic
equations describing Newtonian self-gravitating BECs at $T=0$. We also recall
the qualitative arguments of Chavanis and Harko \cite{chavharko} giving the
scaling of the
maximum mass, minimum radius, and maximum density of relativistic
self-gravitating BECs. In Sec.
\ref{sec_grbec}, we determine the maximum mass of general relativistic BECs
using a partially-relativistic model and compare the result with the ones
obtained by Chavanis and Harko \cite{chavharko} using a semi-relativistic model
and a fully-relativistic model. We also discuss the analogies and the
differences between models that treat neutron stars as fermion stars or as BEC
stars. We finally point out the analogy between  BEC stars described by a stiff
equation of state and the concept of baryon stars introduced by Zel'dovich
\cite{zeldovich}. In Sec. \ref{sec_becc}, we develop a cosmological model in
which dark matter is made of BECs with a stiff equation of state. We point out
the analogy with the model of Zel'dovich \cite{zeldocosmo} that assumes that the
primordial universe is filled with cold baryons. We provide analytical
solutions of the Friedmann equations exhibiting a stiff matter era. We also
discuss the effect of the BEC
equation of state on the evolution of the universe.

\section{Self-gravitating Bose-Einstein condensates}
\label{sec_sgbec}

\subsection{The Gross-Pitaevskii-Poisson system}
\label{sec_gpp}

At $T=0$, in the Newtonian regime, a self-gravitating BEC with short-range interactions is described by the Gross-Pitaevskii-Poisson system
\begin{eqnarray}
\label{gen1}
i\hbar \frac{\partial\psi}{\partial t}=-\frac{\hbar^2}{2m}\Delta\psi+m\Phi\psi+\frac{4\pi a_s\hbar^2N}{m}|\psi|^2\psi,
\end{eqnarray}
\begin{equation}
\Delta\Phi=4\pi G N m |\psi|^2,  \label{gen2}
\end{equation}
where $\rho({\bf r},t)=N m|\psi|^2$ is the mass density ($N$ is the number of
bosons and $m$ is their mass), $\psi({\bf r},t)$ is the wave function,
$\Phi({\bf r},t)$ is the gravitational potential, and $a_s$ is the s-scattering
length of the bosons. These equations are valid in a mean field approximation
which is known to be exact for systems with long-range interactions (such as
self-gravitating systems) when $N\rightarrow +\infty$.

Using the Madelung \cite{madelung} transformation
\begin{eqnarray}
\label{gen3}
\psi=\sqrt{\frac{\rho}{Nm}}e^{iS/\hbar},\qquad {\bf u}=\frac{1}{m}\nabla S,
\end{eqnarray}
where $S({\bf r},t)$ is an action and ${\bf u}({\bf r},t)$ is an irrotational velocity field, we can rewrite the GPP system (\ref{gen1})-(\ref{gen2}) in the form of hydrodynamic equations
\begin{eqnarray}
\label{gen4}
\frac{\partial\rho}{\partial t}+\nabla\cdot (\rho {\bf u})=0,
\end{eqnarray}
\begin{eqnarray}
\label{gen5}
\frac{\partial {\bf u}}{\partial t}+({\bf u}\cdot \nabla){\bf u}=-\frac{1}{\rho}\nabla P-\nabla\Phi-\frac{1}{m}\nabla Q,
\end{eqnarray}
\begin{equation}
\Delta\Phi=4\pi G \rho,  \label{gen6}
\end{equation}
where
\begin{eqnarray}
\label{gen7}
Q=-\frac{\hbar^2}{2m}\frac{\Delta\sqrt{\rho}}{\sqrt{\rho}}
\end{eqnarray}
is the quantum potential and
\begin{eqnarray}
\label{gen8}
P=\frac{2\pi a_s\hbar^2}{m^3}\rho^2
\end{eqnarray}
is the pressure arising from the short-range interaction. It corresponds to a polytropic equation of state
\begin{eqnarray}
\label{gen9}
P=K\rho^{\gamma},\qquad \gamma=1+\frac{1}{n},
\end{eqnarray}
with a polytropic index $n=1$ (i.e. $\gamma=2$) and a polytropic constant
\begin{equation}
\label{gen9b}
K=\frac{2\pi \hbar ^{2}a_s}{m^{3}}.
\end{equation}
Eqs. (\ref{gen4})-(\ref{gen7}) form the quantum barotropic Euler-Poisson
system.

The condition of hydrostatic equilibrium ($\partial_t=0$ and ${\bf u}={\bf 0}$) writes
\begin{eqnarray}
\label{gen10}
\nabla P+\rho\nabla\Phi+\frac{\rho}{m}\nabla Q={\bf 0}.
\end{eqnarray}
It expresses the balance between the gravitational attraction and the repulsion
due to the scattering and to the quantum pressure. Combining this equation with
the Poisson equation (\ref{gen6}) and using Eqs. (\ref{gen7}) and (\ref{gen8}),
we get
\begin{eqnarray}
\label{gen11}
\frac{4\pi a_s\hbar^2}{m^3}\Delta\rho+4\pi G\rho-\frac{\hbar^2}{2m^2}\Delta\left (\frac{\Delta\sqrt{\rho}}{\sqrt{\rho}}\right )=0.
\end{eqnarray}
This differential equation determines the density profile of a self-gravitating
BEC. It is equivalent to the stationary solution (soliton) of the GPP system 
\cite{prd1,bookspringer}. It has been solved analytically (approximately) and
numerically (exactly) in Refs. \cite{prd1} and \cite{prd2} for arbitrary values
of the scattering length $a_s$ and of the boson mass $m$.

\subsection{The Thomas-Fermi approximation}
\label{sec_tf}

In the TF approximation valid when $G M^2 m a_s/\hbar^2\gg 1$, we can neglect the contribution of the quantum potential. In that case, the condition of hydrostatic equilibrium reduces to the usual form
\begin{eqnarray}
\label{tf1}
\nabla P+\rho\nabla\Phi={\bf 0},
\end{eqnarray}
and the differential equation (\ref{gen11}) becomes
\begin{eqnarray}
\label{tf2}
\frac{4\pi a_s\hbar^2}{m^3}\Delta\rho+4\pi G\rho=0.
\end{eqnarray}
Writing $\rho=\rho_0\theta$ and $r=(a_s\hbar^2/Gm^3)^{1/2}\xi$, where $\rho_0$ is the central density, and considering a spherically symmetric system, this equation can be put in the form of the Lane-Emden equation
\begin{equation}
\label{tf3}
\frac{1}{\xi ^{2}}\frac{d}{d\xi }\left (\xi ^{2}\frac{d\theta }{d\xi }\right )=-\theta,
\end{equation}
\begin{equation}
\label{tf4}
\theta(0)=1,\qquad \theta'(0)=0,
\end{equation}
for a polytrope of index $n=1$ \cite{chandra}. It has the analytical solution
\begin{equation}
\label{tf5}
\theta \left( \xi \right) =\frac{\sin \xi }{\xi }.
\end{equation}
The radius of the configuration is defined by the condition $\theta \left( \xi
_{1}\right) =0$, giving $\xi _1=\pi $. Therefore the radius $R$ of the self-gravitating BEC
is given by
\begin{equation}
\label{tf6}
R=\pi \sqrt{\frac{a_s\hbar ^{2}}{Gm^{3}}}.
\end{equation}
It is independent on the central density and on the mass of the system, and depends only on the physical
characteristics  of the condensate (the mass $m$ and the scattering length $a_s$ of the bosons). Actually, it is fixed by the ratio $a_s/m^3$.

The mass of a self-gravitating BEC star with a quartic non-linearity is given as
a function of the
central density and of the coherent scattering length $a_s$ by
\begin{equation}
\label{tf7}
M=4\pi \left( \frac{a_s\hbar
^{2}}{Gm^{3}}\right) ^{3/2}\rho_{0}\xi _{1}^{2}\left| \theta ^{\prime }\left( \xi _{1}\right)
\right| ,
\end{equation}
yielding
\begin{equation}
\label{tf8}
M=4\pi ^{2}\left( \frac{a_s\hbar
^{2}}{Gm^{3}}\right) ^{3/2}\rho _{0},
\end{equation}
where we have used $\left| \theta ^{\prime }\left( \xi _{1}\right)
\right| =1/\pi $. Using Eq. (\ref{tf6}), it can be expressed in terms of the radius
and central density by
\begin{equation}
M=\frac{4}{\pi }\rho _{0}R^{3},
\label{tf9}
\end{equation}
which shows that the mean density of the configuration $\overline{\rho}=3M/4\pi R^{3}$ is related to the central density by the relation $\overline{\rho}=3\rho _{0}/\pi ^{2}$. Other quantities of interest such as the energy and the moment of inertia are derived in \cite{prd1}.

\subsection{Maximum mass of relativistic BEC stars: qualitative treatment and fundamental scalings}
\label{sec_grz}

The Newtonian treatment of self-gravitating BECs is appropriate to describe dark matter halos. However, general relativistic effects may be important in the case of BEC stars describing compact objects such as neutron stars or dark matter stars \cite{bookspringer}.

The radius of a Newtonian BEC star is given by Eq. (\ref{tf6}). In the Newtonian
treatment, there is no limit on the mass of the BEC. However, the Newtonian
treatment breaks down when the radius of the star approaches the Schwarzschild
radius $R_S=2GM/c^2$. Equating the two radii, namely writing $M={Rc^2}/{2G}$
with $R$ given by Eq. (\ref{tf6}), and ignoring the prefactors that are
necessarily inexact, we obtain the scaling of the maximum mass and of the
minimum radius of a relativistic BEC star \cite{chavharko}:
\begin{equation}
{M_*}=\frac{\hbar c^2\sqrt{a_s}}{(Gm)^{3/2}}=1.420\, \kappa \, {M_{\odot}},
\label{grz1}
\end{equation}
\begin{equation}
R_*=\frac{GM_*}{c^2}=\left (\frac{a_s\hbar^2}{Gm^3}\right )^{1/2}=2.106\, \kappa\;{\rm km},
\label{grz2}
\end{equation}
where we have introduced the dimensionless parameter
\begin{equation}\label{kappa}
\kappa=\left(\frac{a_s}{1\;{\rm fm}}\right)^{1/2}\left(\frac{m}{2m_n}\right)^{-3/2}.
\end{equation}
From Eqs. (\ref{grz1}) and (\ref{grz2}), we obtain the scaling of the maximum  central density
\begin{equation}
\rho_*=\frac{m^3c^2}{2\pi a_s\hbar^2}=4.846\times 10^{16}\, \kappa ^{-2}\;{\rm g/cm^3},
\label{grz3}
\end{equation}
where the factor $2\pi$ has been introduced for future convenience.

We note that the expression of the scaled radius $R_*$ is the same as in
the Newtonian regime (it is independent on $c$) while the scaling of the mass
and of the density are determined by relativistic effects.

\section{General relativistic Bose-Einstein condensate stars}
\label{sec_grbec}

For a correct determination
of the maximum mass of BEC stars, we cannot ignore
the effects induced by the space-time curvature, and a general relativistic
treatment is necessary.

\subsection{The Tolman-Oppenheimer-Volkoff equation}
\label{sec_tov}

For a static spherically symmetric star, the interior line element is given by
\begin{equation}
\label{tov0}
ds^{2}=e^{\nu (r)}c^2dt^{2}-e^{\lambda (r)}dr^{2}-r^{2}\left(
d\theta ^{2}+\sin ^{2}\theta d\phi ^{2}\right).
\end{equation}
The equations describing a general relativistic compact
star are the mass continuity equation and the
Tolman-Oppenheimer-Volkoff (TOV) equation. They write \citep{Gl00}:
\begin{equation}
\frac{dM}{dr}=4\pi \frac{\epsilon}{c^2} r^2 , \label{tov1}
\end{equation}
\begin{equation}
\frac{dP}{dr}=-\frac{G \left(\epsilon+P\right)
\left\lbrack 4\pi Pr^{3}/c^{2}+M(r)\right\rbrack }{r^{2}c^2\left[ 1-2GM(r)/c^{2}r\right] },  \label{tov2}
\end{equation}
where $\epsilon$ is the energy density and $M(r)$ is the total mass interior to
$r$. The mass of the star is $M=M(R)$ where $R$ is its radius. These equations
extend the classical condition of hydrostatic
equilibrium for a self-gravitating gas to the context of general
relativity. The system of equations (\ref{tov1})-(\ref{tov2}) must be closed by
choosing the equation of state $P=P(\epsilon)$ for the thermodynamic pressure.
At the center of the star, the mass must satisfy the boundary
condition $M(0)=0$. For the thermodynamic pressure $P$, we assume that it vanishes on
the surface: $P(R)=0$.

The exterior of the star is
characterized
by the Schwarzschild metric, describing the vacuum ($P=\epsilon=0$) outside the
star, and given by \citep{Gl00}:
\begin{equation}
\left( e^{\nu }\right) ^{\rm ext}=\left( e^{-\lambda }\right) ^{\rm
ext}=1-\frac{2GM}{c^2r},\qquad r\geq R.
\end{equation}
The interior solution must match with the exterior solution on the
vacuum boundary of the star.

The components of the metric tensor are determined by
\begin{equation}
e^{-\lambda(r)}=1-\frac{2GM(r)}{r c^2},
\label{tov2a}
\end{equation}
\begin{equation}
\frac{dP}{dr}+\frac{P+\epsilon}{2}\frac{d\nu}{dr}=0,\qquad
e^{\nu(R)}=1-\frac{2GM}{Rc^2}.
\label{tov2b}
\end{equation}
The boundary condition on $e^{\nu}$ has been chosen so that this component is
continuous with the exterior solution at $r=R$.

\subsection{Maximum mass of relativistic BEC stars with short-range interactions: partially-relativistic treatment}
\label{sec_mm}

We consider a partially-relativistic model (see Appendix \ref{sec_partially}) in which the BEC star is described in general relativity by the  equation of state
\begin{equation}
\label{mm1}
P=K\rho^2, \qquad \epsilon=\rho c^2+P,
\end{equation}
where $K$ is given by Eq. (\ref{gen9b}). Here, $\epsilon$ is the energy density
and $\rho$ is the rest-mass density. It is related to the number density $n$ by
$\rho=m n$. The pressure can
be expressed as a function of the energy density as (see Appendix
\ref{sec_partially}):
\begin{equation}
P=\frac{c^4}{4K}\left (\sqrt{1+\frac{4K\epsilon}{c^4}}-1\right )^2.
\label{mm2}
\end{equation}
In the non-relativistic regime ($\epsilon\rightarrow 0$), we recover the classical equation of state of a BEC star $P\sim K\epsilon^2/c^4\sim K\rho^2$. In the ultra-relativistic regime ($\epsilon\rightarrow +\infty$), we obtain a stiff equation of state $P\sim \epsilon$ in which the velocity of sound is equal to the velocity of light. A stiff equation of state was first introduced by Zel'dovich \cite{zeldovich} in the context of baryon stars in which the baryons interact through a vector meson field (see Sec. \ref{sec_baryon}). We know that a linear equation of state $P=q\epsilon$ leads to a mass-central density
relation that presents damped oscillations, and to a mass-radius relation that
has a spiral structure \cite{chavgen,chavbh}.  Therefore, the series of
equilibria of BEC stars described by the equation of state (\ref{mm2}) will
exhibit this behavior. This is similar to the series of equilibria of neutron
stars modeled as a gas of relativistic fermions that have a linear equation of
state $P\sim \epsilon/3$ for $\epsilon\rightarrow +\infty$ (see Sec.
\ref{sec_mma}) \cite{ov,misner,mt}. This is also similar to the series of
equilibria of isothermal spheres described by a linear equation of state $P=\rho
k_B T/m$ in Newtonian gravity \cite{aaiso}.

The equation of state (\ref{mm1}) is a particular case, corresponding to a polytropic index $n=1$, of the class of equations of state studied by Tooper \cite{tooper2} in general relativity. We shall use his formalism and notations. Therefore, we set
\begin{equation}\label{mm4}
\rho =\rho_0 \theta,\quad r=\frac{\xi}{A}, \quad \sigma=\frac{K\rho_0}{c^2},
\end{equation}
\begin{equation}\label{mm5}
M(r)=\frac{4\pi\rho_0}{A^3}v(\xi), \quad A=\left (\frac{2\pi G}{K}\right )^{1/2},
\end{equation}
where $\rho_0$ is the central rest-mass density and $\sigma$ is the relativity
parameter.  In terms of these variables,
the TOV equation and the mass continuity equation become
\begin{equation}\label{mm6}
\frac{d\theta}{d\xi}=-\frac{(1+2\sigma\theta)(v+\sigma\xi^3\theta^2)}{\xi^2(1-4\sigma v/\xi)},
\end{equation}
\begin{equation}\label{mm7}
\frac{dv}{d\xi}=\theta \xi^2 (1+\sigma\theta).
\end{equation}
For a given value of the relativity parameter $\sigma$, they have to be solved with the initial condition $\theta(0)=1$ and $v(0)=0$. Since $v\sim \xi^3$ as $\xi\rightarrow 0$, it is clear that  $\theta'(0)=0$. On the other hand, the density vanishes at the first zero $\xi_1$ of $\theta$: $\theta(\xi_1)=0$.
This determines the boundary of the star. In the non-relativistic limit $\sigma\rightarrow 0$, the system of equations  (\ref{mm6})-(\ref{mm7}) reduces to the Lane-Emden equation (\ref{tf3}) with $n=1$.

From the foregoing relations, we find that the radius, the mass and the central density  of the configuration are given by
\begin{equation}\label{mm9}
R=\xi_1R_*, \quad M=2\sigma v(\xi_1)M_*, \quad \rho_{0}=\sigma \rho_*,
\end{equation}
where
\begin{equation}\label{mm9b}
R_*=\left (\frac{K}{2\pi G}\right )^{1/2},\quad M_*=\left (\frac{Kc^4}{2\pi G^3}\right )^{1/2},\quad \rho_*=\frac{c^2}{K}.
\end{equation}
For the value of $K$ given by Eq. (\ref{gen9b}), one can check that the fundamental scaling parameters $R_*$, $M_*$ and $\rho _*$ are given by Eqs. (\ref{grz1})-(\ref{grz3}). By varying $\sigma$ from $0$ to $+\infty$, we obtain the series of equilibria in the form $M(\rho_0)$ and $R(\rho_0)$. We can then plot the mass-radius relation $M(R)$ parameterized by $\rho_0$.

Using the Poincar\'e theorem \cite{poincare} (see also \cite{katzpoincare,ijmpb}), one can
show \cite{chavgen,chavbh} that the series of equilibria becomes unstable after the first
mass peak and that a new mode of instability appears at each turning
point of mass in the series of equilibria (see \cite{st} for an alternative
derivation of these results based on the equation of pulsations). These results
of dynamical stability for general
relativistic stars are similar to results of dynamical and
thermodynamical stability for Newtonian self-gravitating systems
\cite{aaiso,aaantonov}.

\begin{figure}[!ht]
\includegraphics[width=0.98\linewidth]{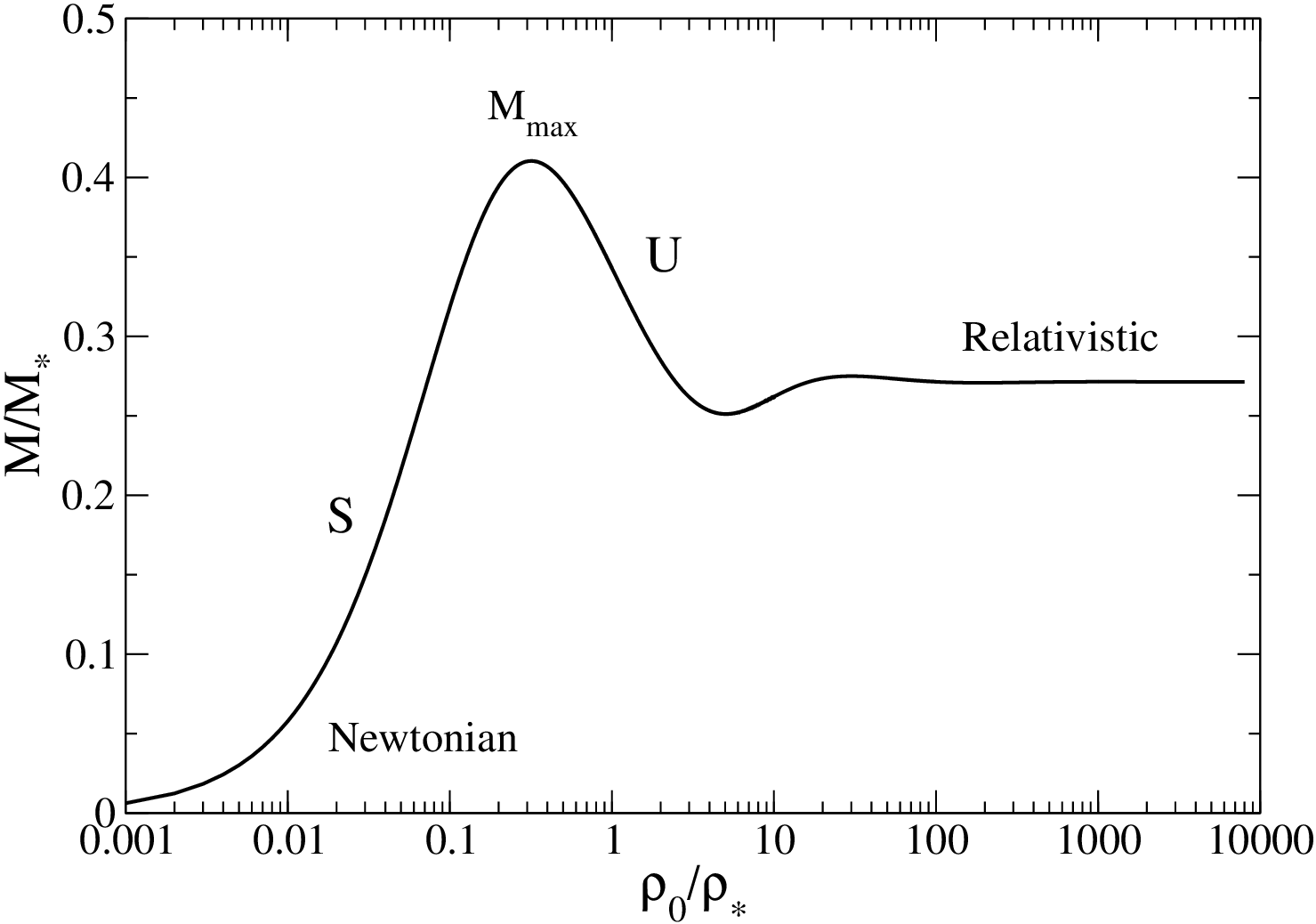}
\caption{Dimensionless
mass-central density relation of a relativistic BEC with short-range
interactions modeled by the equation of state (\ref{mm1}). There exist a maximum
mass $M_{{\rm max}}/M_{*}=0.4104$ at which the series of equilibria becomes
dynamically unstable. The velocity of sound is always smaller than the velocity
of light. We note that the  mass-central density relation presents damped
oscillations at high densities similarly to neutron stars described by a
fermionic equation of state \cite{ov,misner,mt,chavgen,chavbh}. \label{rhoM}}
\end{figure}

\begin{figure}[!ht]
\includegraphics[width=0.98\linewidth]{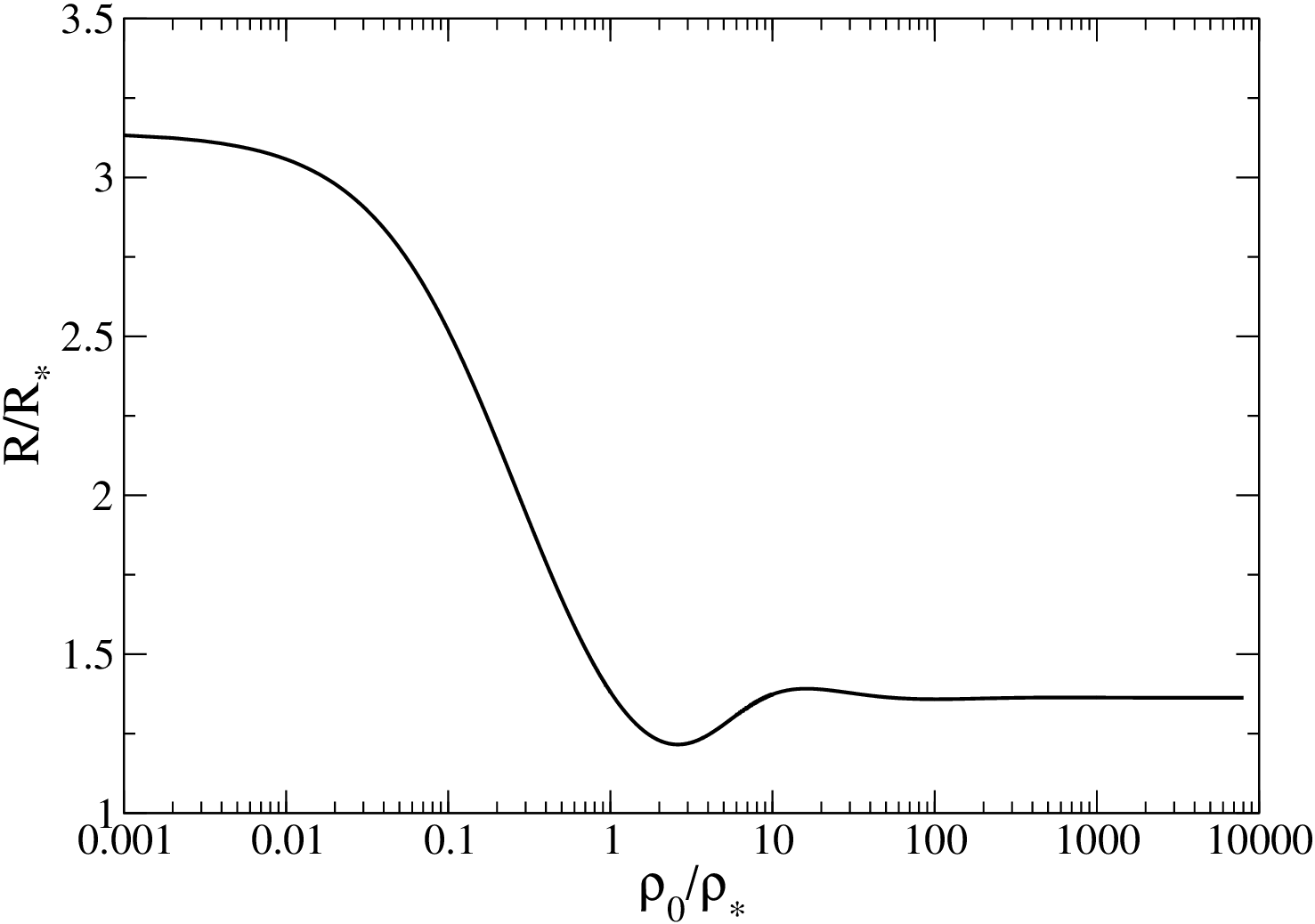}
\caption{ Dimensionless
radius-central density relation of a relativistic BEC with short-range
interactions modeled by the equation of state (\ref{mm1}). \label{rhoR}}
\end{figure}

\begin{figure}[!ht]
\includegraphics[width=0.98\linewidth]{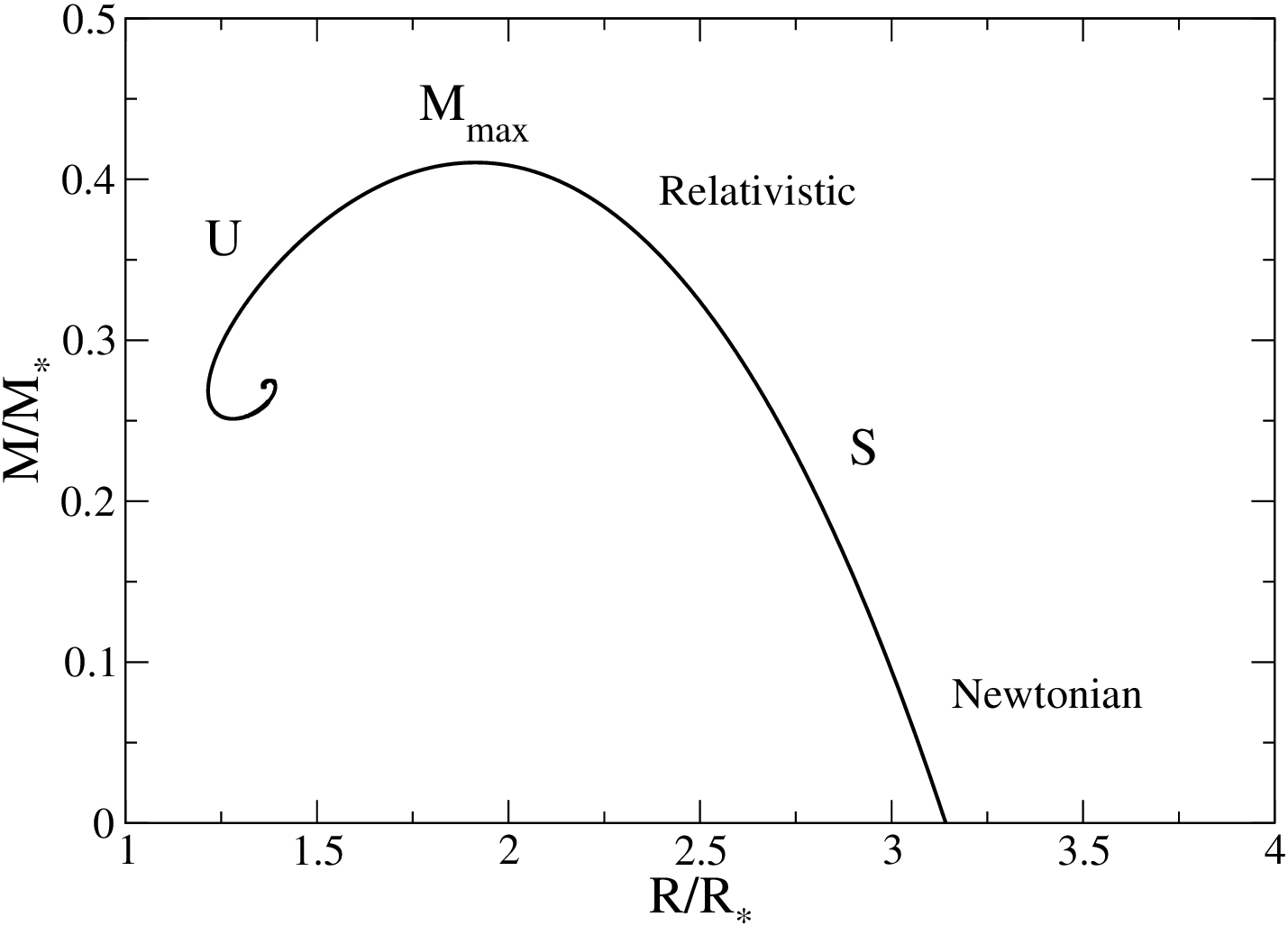}
\caption{Dimensionless mass-radius relation of a relativistic BEC with
short-range interactions modeled by the equation of state (\ref{mm1}). The
series of equilibria is parameterized by the relativity parameter $\sigma$. The
mass-radius relation presents a snail-like structure (spiral) at high densities
similarly to neutron stars described by a fermionic equation of state
\cite{ov,misner,mt,chavgen,chavbh}. There exist a maximum mass $M_{{\rm
max}}/M_*=0.4104$ and a minimum radius $R_{{\rm min}}/R_*=1.914$ corresponding
to a maximum central density $(\rho_0)_{{\rm max}}=0.318\rho_*$. There also
exist a maximum radius $R_{{\rm max}}/R_*=\pi$ corresponding to the Newtonian
limit $\sigma\rightarrow 0$. \label{RM}}
\end{figure}

\begin{figure}[!ht]
\includegraphics[width=0.98\linewidth]{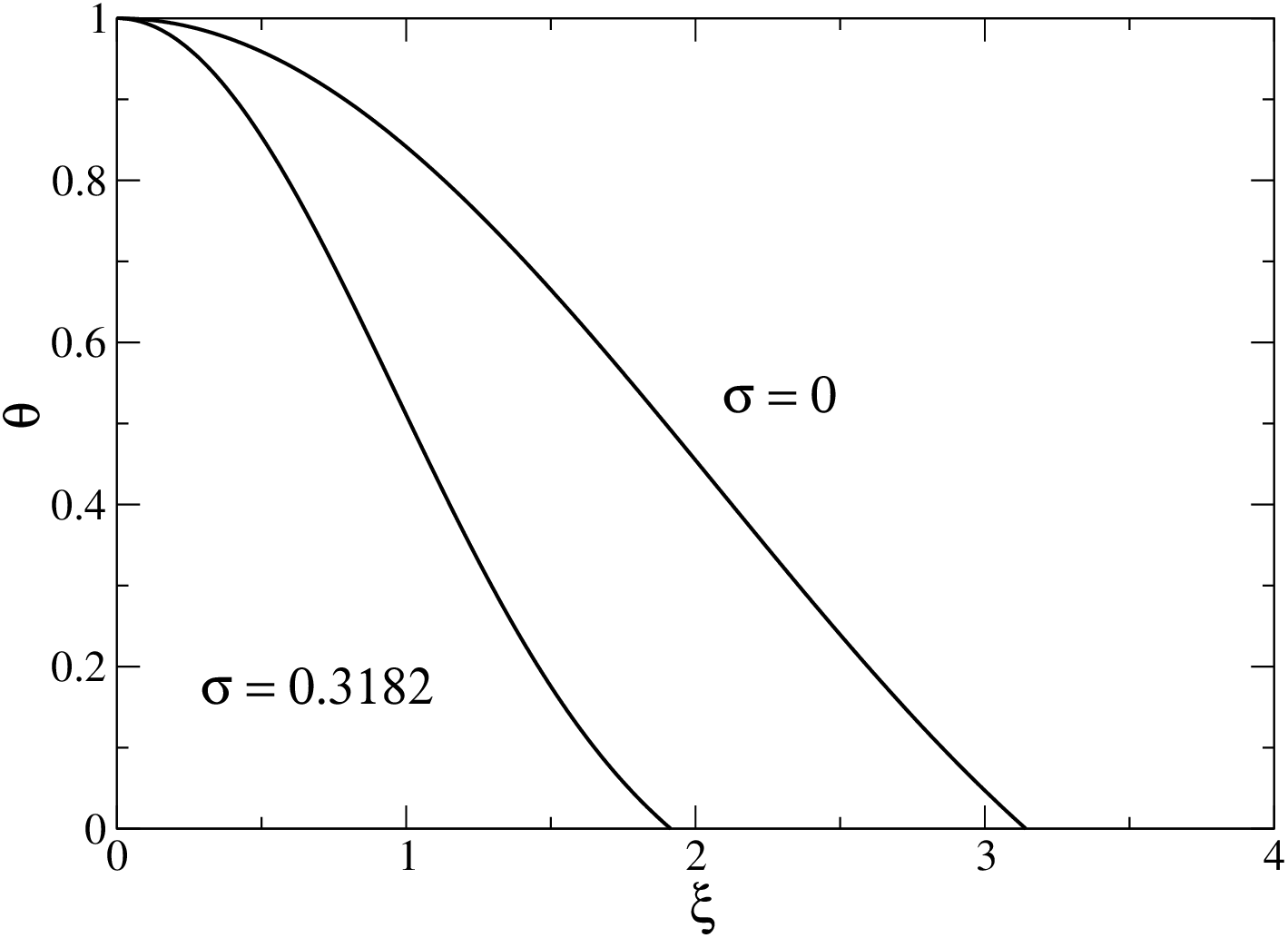}
\caption{Dimensionless density profiles corresponding to $\sigma=0$ (Newtonian) and $\sigma=\sigma_c=0.318$ (maximum mass).\label{prof}}
\end{figure}

The series of equilibria corresponding to the equation of state (\ref{mm1}) is represented in Figs. \ref{rhoM}-\ref{RM}. These figures respectively give the mass-central density relation, the radius-central density relation, and the mass-radius relation. Some density profiles are plotted in Fig. \ref{prof}. The series of equilibria is parameterized by the relativity parameter $\sigma$ going from $\sigma=0$ (non-relativistic) to $\sigma\rightarrow +\infty$ (ultra-relativistic). The configurations are stable for $\sigma\le \sigma_{c}$ and unstable for
 $\sigma\ge \sigma_{c}$ where
\begin{equation}\label{mm14}
\sigma_{c}=0.318
\end{equation}
corresponds to the first turning point of mass: $M'(\sigma_c)=0$. The values of $\xi_1$ and $v(\xi_1)$ at this point are
\begin{equation}\label{mm15}
\xi_1=1.914,\qquad v(\xi_1)=0.6453.
\end{equation}
The corresponding values of radius, mass and central density are
\begin{equation}\label{mm16}
R_{{\rm min}}=1.914\left (\frac{a_s\hbar^2}{Gm^3}\right )^{1/2}=4.03\, \kappa
\;{\rm km},
\end{equation}
\begin{equation}\label{mm17}
{M_{{\rm max}}}=0.4104\frac{\hbar c^2\sqrt{a_s}}{(Gm)^{3/2}}=0.583\, \kappa\,
{M_{\odot}},
\end{equation}
\begin{equation}\label{mm18}
(\rho_0)_{{\rm max}}=0.318\frac{m^3c^2}{2\pi a_s\hbar^2}=1.54\times 10^{16}\,
\kappa ^{-2}\;{\rm g/cm^3},
\end{equation}
respectively. They define the minimum radius, the maximum mass, and the maximum
rest-mass density of the stable configurations.

The energy density is related to the rest-mass density by Eq. (\ref{mm1}) which can be rewritten as
\begin{equation}\label{mm18b}
\epsilon/c^2=\rho \left (1+\rho \frac{2\pi \hbar^2 a_s}{m^3c^2}\right ).
\end{equation}
Using Eqs. (\ref{mm18}) and (\ref{mm18b}), the maximum energy density is
\begin{equation}\label{mm18c}
(\epsilon_0)_{{\rm max}}/c^2=0.419\frac{m^3c^2}{2\pi a_s\hbar^2}=2.03\times 10^{16}\,
\kappa ^{-2}\;{\rm g/cm^3}.
\end{equation}

We note that the radius of a relativistic BEC star is
necessarily smaller than
\begin{equation}\label{mm19}
R_{\rm max}=\pi \sqrt{\frac{\hbar ^{2}a_s}{Gm^{3}}}=6.61\, \kappa \, {\rm km},
\end{equation}
corresponding to the Newtonian limit ($\sigma\rightarrow 0$). The Newtonian
approximation is valid for small masses $M\ll M_{max}$. The radius decreases as
$M$ increases until the maximum mass and the minimum radius are reached. When
$M>M_{max}$, there is no equilibrium state and the BEC star is expected to
collapse and form a black hole. When $M<M_{max}$, there exist stable
equilibrium states with $R_{min}<R<R_{max}$ that correspond to BEC stars for
which
gravitational collapse is prevented by quantum mechanics (the self-interaction
of the bosons). We note that the radius of the BEC star
is very much constrained as it lies in the range $4.03\, \kappa\le R ({\rm km}) \le 6.61\, \kappa$.

A quantity of physical interest is the mass-radius ratio
\begin{equation}\label{mr1}
\frac{2GM}{Rc^2}=\frac{4\sigma v(\xi_1)}{\xi_1}.
\end{equation}
At the critical point, the value of the mass-radius ratio is $0.429$. We check
that it is smaller than the Buchdahl maximum bound $2GM/Rc^2=8/9=0.888$
corresponding to constant density stars \cite{buchdahl}.

\subsection{Comparison between the different models}
\label{sec_comp}

The values of the maximum mass, minimum radius, and maximum central energy
density of general relativistic BEC stars can be written as
\begin{equation}\label{comp1}
R_{{\rm min}}=A_1\left (\frac{a_s\hbar^2}{Gm^3}\right )^{1/2}=A_1'\, \kappa
\;{\rm km},
\end{equation}
\begin{equation}\label{comp2}
{M_{{\rm max}}}=A_2\frac{\hbar c^2\sqrt{a_s}}{(Gm)^{3/2}}=A_2'\, \kappa\,
{M_{\odot}},
\end{equation}
\begin{equation}\label{comp3}
(\epsilon_0)_{{\rm max}}/c^2=A_3\frac{m^3c^2}{2\pi a_s\hbar^2}=A_3' \times 10^{16}\,
\kappa ^{-2}\;{\rm g/cm^3},
\end{equation}
where $\kappa$ is defined by Eq. (\ref{kappa}). These scalings are  fundamental
for BEC stars \cite{chavharko}. However, the values of the prefactors depend on
the relativistic model.

The best model is the one based on the equation of state (\ref{fully1})
considered in Sec. VI.C. of Chavanis and Harko \cite{chavharko} because this
equation of state can be derived from the relativistic Klein-Gordon equation
\cite{colpi}. Therefore, this model is fully-relativistic, both regarding the
equation of state and the treatment of gravity. In that model, the prefactors
are $A_1=1.923$, $A_1'=4.047$, $A_2=0.307$, $A_2'=0.436$, $A_3=0.398$, and
$A_3'=1.929$. They can be considered as being the exact prefactors for
relativistic BEC stars.

The model based on the equation of state (\ref{semi1}) considered in Sec.
VI.B. of Chavanis \& Harko \cite{chavharko} is very approximate because it is
based on an equation of state $P=2\pi \hbar ^{2}a_s\rho^2/m^{3}$ derived from
the classical GP equation and it furthermore assumes that the energy density is
dominated by the rest-mass density so that $\epsilon=\rho c^2$. Therefore, this
model is semi-relativistic because the equation of state is classical while
gravity is treated in the framework of general relativity. In that model, the
prefactors are $A_1=1.888$, $A_1'=3.974$,  $A_2=0.5001$, $A_2'=0.710$,
$A_3=0.42$, and $A_3'=2.035$.

The model based on the equation of state (\ref{partially4})  is
intermediate between the two previous models. It is based on  an equation of
state $P=2\pi \hbar ^{2}a_s\rho^2/m^{3}$  derived from the classical GP equation
but it takes into account the difference between the energy density and the
rest-mass density due to pressure effects: $\epsilon=\rho c^2+P$
\cite{tooper2}. Therefore,
this model is partially-relativistic. In that model, the prefactors are
$A_1=1.914$, $A_1'=4.03$, $A_2=0.4104$, $A_2'=0.583$, $A_3=0.419$, and
$A_3'=2.03$.

\begin{figure}[!ht]
\includegraphics[width=0.98\linewidth]{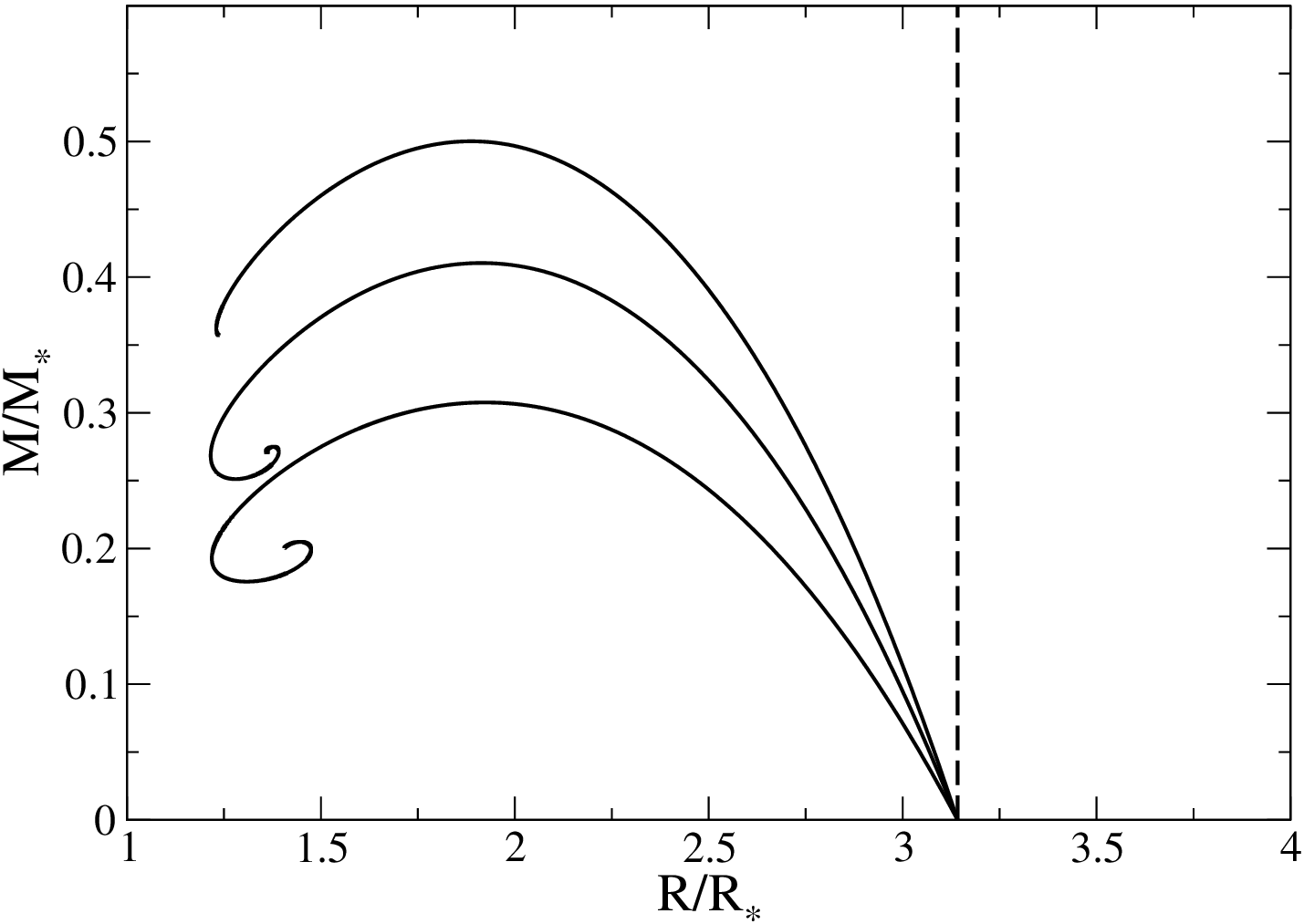}
\caption{Comparison between the mass-radius relations corresponding to the
fully-relativistic model (lower curve), to the partially-relativistic model
(intermediate curve), and to the semi-relativistic model (upper curve). The
vertical dashed line corresponds to the Newtonian limit. \label{newRMtotal}}
\end{figure}

The mass-radius relation of general relativistic BEC stars at $T=0$
corresponding to these different models is plotted in Fig. \ref{newRMtotal}. We
note that the values of the prefactors do not differ much from one model to the
other. The maximum mass varies between $\sim 0.3 M_*$ and $\sim 0.5 M_*$ while
the minimum radius and the maximum energy density almost do not change.

\begin{figure}[!ht]
\includegraphics[width=0.98\linewidth]{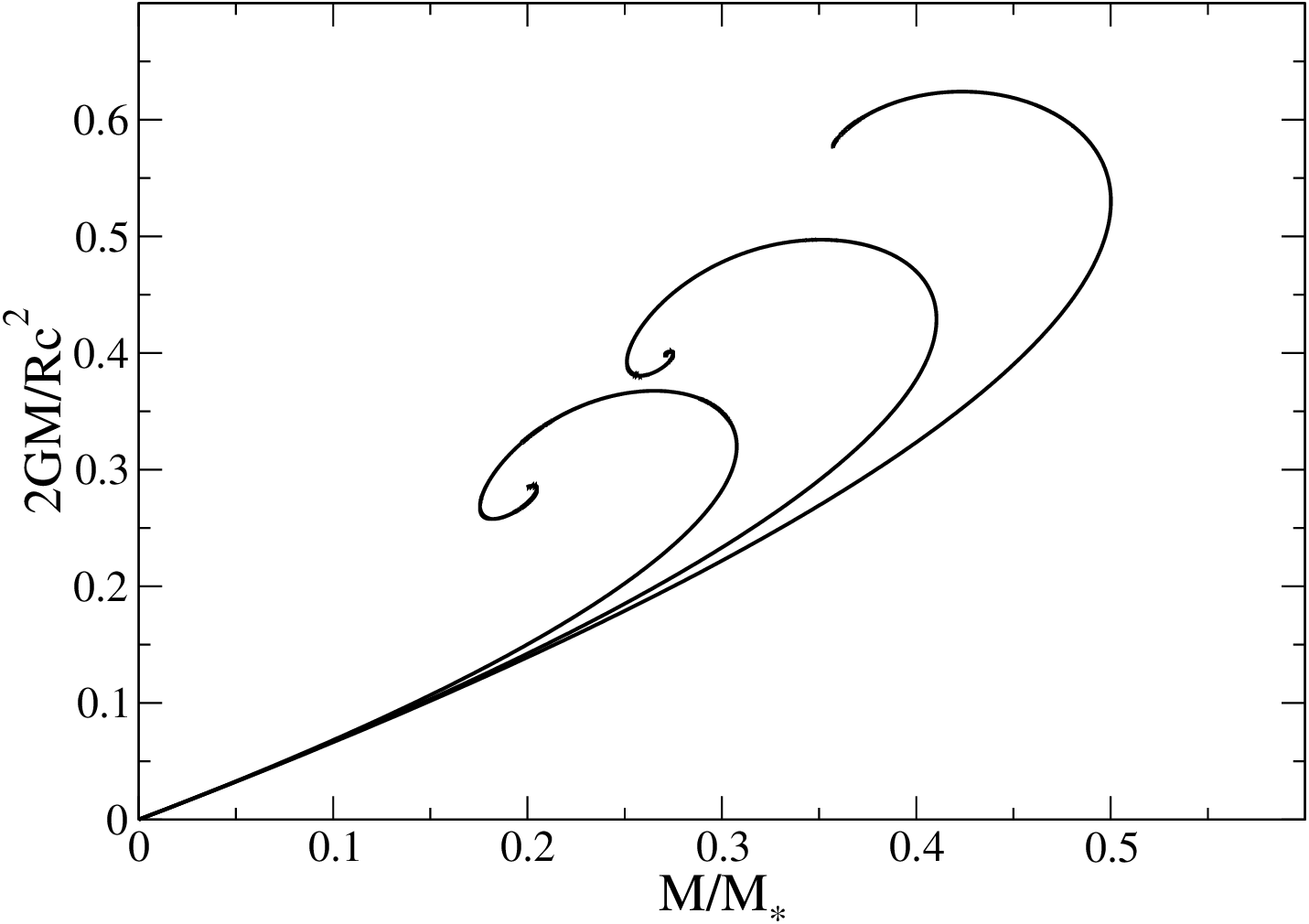}
\caption{Mass-radius ratio of general relativistic BEC stars corresponding to
the fully-relativistic model (left curve), to the partially-relativistic model
(middle curve), and to the semi-relativistic model (right curve). The Buchdahl
maximum bound $2GM/Rc^2=8/9=0.888$ is much higher. \label{newMsRtotal}}
\end{figure}

\begin{table}
\caption[]{Observational values of the mass and radius of neutron stars.}
         \label{table}
      \[
         \begin{array}{l l l l l l l }
            \hline
            \noalign{\smallskip}
{\rm Ref.}    & M/M_{\odot}  & R/{\rm km} & 2GM/Rc^2 \\
            \noalign{\smallskip}
            \hline
            \noalign{\smallskip}
            \lbrack 113\rbrack  & 1.3  & 8 & 0.479  \\
            \lbrack 113\rbrack & 1.95 & 12 & 0.479 \\
	    \lbrack 114\rbrack & 2.1 & 12.5 & 0.4956 \\
	    \lbrack 115\rbrack & 1.58 & 9.11  & 0.512  \\
            \lbrack 116\rbrack & 1.9 & 15.2  & 0.369  \\
           \noalign{\smallskip}
            \hline
         \end{array}
     \]
   \end{table}

As discussed specifically in Sec. \ref{sec_mma}, general relativistic BEC stars
may describe neutron stars with a superfluid core. This is why we have
normalized the mass of the bosons by $2m_n$ (Cooper pair) in Eq. (\ref{kappa}).
However, general relativistic BEC stars may describe other compact  objects such
as boson stars or dark matter stars \cite{bookspringer}. In this respect, it may
be convenient to write the maximum mass, the minimum radius, and the maximum
central density as\footnote{We note that $R_{min}$ and $(\epsilon_0)_{max}$
depend only on $M_{max}$ as a result of relativity. Furthermore, $M_{max}$
depends only on $m$ and $a_s$ through the ratio
$\kappa\propto a_s^{1/2}/m^{3/2}$. Therefore, there is only one free parameter
$\kappa$ in the theory.} 
\begin{equation}
\frac{M_{\rm max}}{M_{\odot}}=A\, \left (\frac{a}{\rm fm}\right )^{1/2}\left (\frac{{\rm GeV}/c^2}{m}\right )^{3/2},
\label{grz4}
\end{equation}
\begin{equation}
\frac{R_{\rm min}}{\rm km}=B\, \frac{M_{\rm max}}{M_{\odot}},
\label{grz5}
\end{equation}
\begin{equation}
\frac{(\epsilon_{0})_{\rm max}/c^2}{ 10^{16}\, {\rm g/cm^3}}=C\, \left
(\frac{M_{\odot}}{M_{\rm max}}\right )^2.
\label{grz4b}
\end{equation}
For the fully-relativistic model, $A=1.12$, $B=9.27$, and $C=0.364$. For the
semi-relativistic model, $A=1.83$, $B=5.59$, and $C=1.02$. For the
partially-relativistic model, $A=1.50$, $B=6.91$, and $C=0.689$. The maximum
radius $R_{\rm max}$ of the star, corresponding to the radius of a Newtonian BEC
given by Eq. (\ref{mm19}), can be written as
\begin{equation}
\frac{R_{\rm max}}{\rm km}=17.0\, \left (\frac{a}{\rm fm}\right )^{1/2}\left (\frac{{\rm GeV}/c^2}{m}\right )^{3/2}.
\label{sreu}
\end{equation}
Finally, the value of the mass-radius ratio $2GM/Rc^2=2.95 (M/M_{\odot})({\rm
km}/R)$ of general
relativistic BEC stars at the critical point is $0.319$ in the
fully-relativistic model, $0.529$ in the semi-relativistic model, and $0.429$ in
the partially-relativistic model. It  varies between $\sim 0.3$ and $\sim 0.5$
depending on the model.  The mass-radius ratio  is plotted as a function of
$M/M_*$ in Fig.
\ref{newMsRtotal} for the different models. We note that the value of
$2GM/Rc^2$ at the critical point provides the maximum value  of the mass-radius
ratio for the stable part of the series of equilibria.

The observations of neutron stars compiled by Mukherjee {\it
et al.} \cite{mukherjee} give a value of the mass-radius ratio  $2GM/Rc^2\sim
0.5$ (see Table \ref{table}). This is
substantially larger than the value $0.319$  predicted from the fully
relativistic equation of
state  (\ref{fully1}). In other words, the predicted radius of the neutron stars
is larger than observed. This led Mukherjee {\it
et al.} \cite{mukherjee} to conclude that the BEC model is ruled out.
However, their conclusion may be too pessimistic because several effects can
alter the equation of state of the BEC. For
example, as they note, the interior of neutron stars could be a composition of
BECs of kaons
or pions. This may change the  mass-radius relation of neutron
stars. On the other hand, we note that the value $0.5$ is relatively
close to the values $0.429$ and $0.529$  obtained from the equations of state 
(\ref{partially4}) and (\ref{semi1}). This agreement is puzzling
because these
equations of state are less-well justified theoretically than the equation of
state  (\ref{fully1}). This may be a further motivation to
study these equations of state, independently of the BEC model. We finally note
that the
value $0.369$ obtained in \cite{straaten} is relatively close to the
value $0.319$ of the fully relativistic BEC model. Additional observations may
be
necessary to reach definite conclusions.

\subsection{On the maximum mass of neutron stars}
\label{sec_mma}

In their seminal paper, Oppenheimer and Volkoff \cite{ov} modeled neutron stars
as a completely degenerate ideal gas of relativistic fermions. In that case,
gravitational collapse is prevented by the Pauli exclusion principle. Since
these objects are very compact, one must use general relativity. Therefore, the
equilibrium configurations of neutron stars in this model are obtained by
solving the TOV equations (\ref{tov1}) and (\ref{tov2}) with the equation of
state $P(\epsilon)$ corresponding to a relativistic fermionic gas at $T=0$. This
equation of state is given in parametric form by \cite{chandra}:
\begin{equation}
P=Af(x),\qquad \rho c^2=8Ax^3,
\label{mma1}
\end{equation}
\begin{equation}
\epsilon=\rho c^2+E_{kin}=8A\left\lbrack x^3+\frac{1}{8}g(x)\right\rbrack,\quad A=\frac{\pi m_n^4c^5}{3h^3},
\label{mma2}
\end{equation}
\begin{equation}
f(x)=x(2x^2-3)(x^2+1)^{1/2}+3 \sinh^{-1} x,
\label{mma3}
\end{equation}
\begin{equation}
g(x)=8x^3\left\lbrack (x^2+1)^{1/2}-1\right\rbrack-f(x).
\label{mma4}
\end{equation}
In the non-relativistic limit ($\epsilon\rightarrow 0$), we get
\begin{eqnarray}
\epsilon\sim \rho c^2&,&\qquad P\sim \frac{1}{5}\left (\frac{3}{8\pi}\right )^{2/3}\frac{h^2}{m_n^{8/3}c^{10/3}}\epsilon^{5/3},\nonumber\\
& &P\sim \frac{1}{5}\left (\frac{3}{8\pi}\right )^{2/3}\frac{h^2}{m_n^{8/3}}\rho^{5/3},
\label{mma5}
\end{eqnarray}
corresponding to a polytrope $n=3/2$. In the ultra-relativistic limit ($\epsilon\rightarrow +\infty$), we get
\begin{eqnarray}
\epsilon\sim \frac{3}{4}\left (\frac{3}{8\pi}\right )^{1/3}\frac{h c}{m_n^{4/3}}\rho^{4/3},\quad P\sim \frac{1}{3}\epsilon,\nonumber\\
P\sim \frac{1}{4}\left (\frac{3}{8\pi}\right )^{1/3}\frac{h c}{m_n^{4/3}}\rho^{4/3},
\label{mma6}
\end{eqnarray}
corresponding to a polytrope $n=3$. In this limit, the equation of state is
linear: $P\sim \epsilon/3$.

\begin{figure}[!ht]
\includegraphics[width=0.98\linewidth]{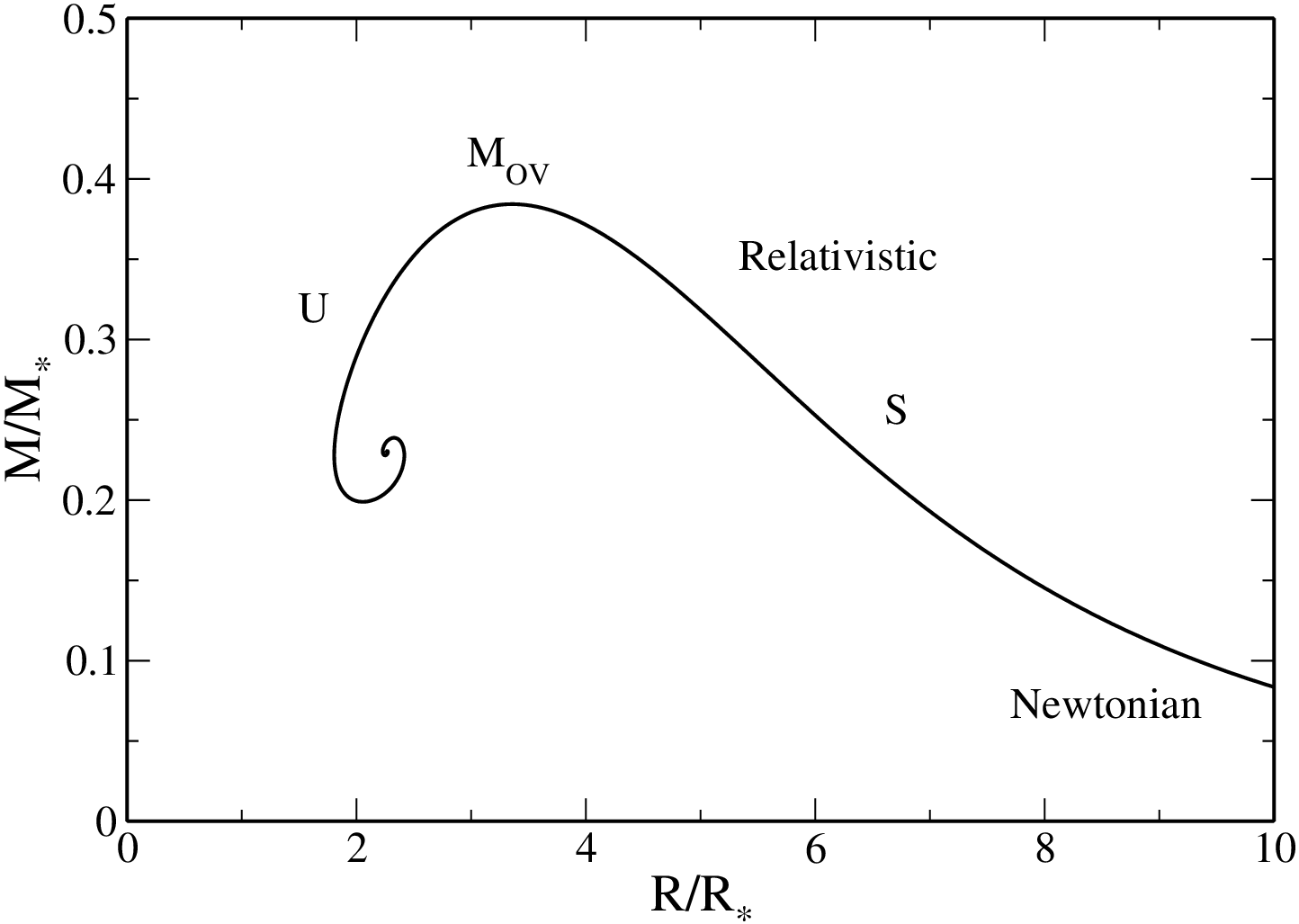}
\caption{Dimensionless mass-radius relation of neutron stars interpreted as
fermion stars (Oppenheimer-Volkoff model) \cite{ov}. There exists a maximum mass
$M_{OV}=0.384 M_*$, a minimum radius $R_{OV}=3.36 R_{*}$, and a maximum central
energy density $\epsilon_{OV}/c^2=2.33\times 10^{-2} \epsilon_*$ with
$M_*=(\hbar c/G)^{3/2}/m_n^2$, $R_{*}=GM_*/c^2=(\hbar^3/Gc)^{1/2}/m_n^2$ and
$\epsilon_*/c^2=M_*/R_*^3=m_n^4 c^3/\hbar^3$. \label{MROV}}
\end{figure}

\begin{figure}[!ht]
\includegraphics[width=0.98\linewidth]{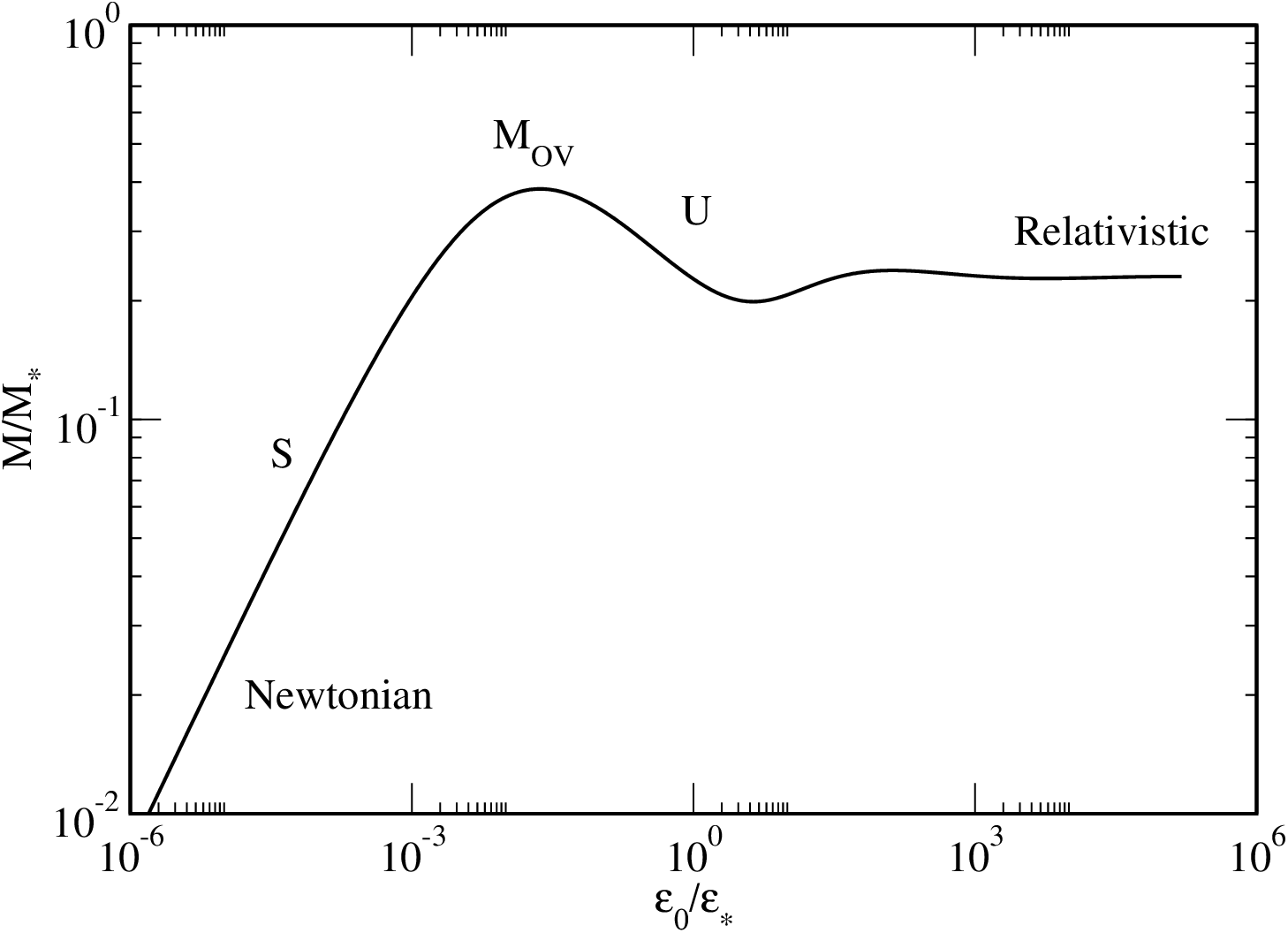}
\caption{Dimensionless mass-central energy density relation of neutron stars
interpreted as fermion stars (Oppenheimer-Volkoff model) \cite{ov}.
\label{epsMOV}}
\end{figure}

The mass-central density relation of fermion stars
presents damped oscillations and the mass-radius relation has a snail-like
(spiral) structure (see Figs. \ref{MROV} and
\ref{epsMOV}) \cite{ov,misner,mt,chavgen,chavbh}. The maximum mass, minimum
radius, and maximum
energy density are\footnote{We note
that  Tooper \cite{tooper2}, in his Sec. IX.b, considers a simplified model of
neutron stars by using the non-relativistic equation of state for fermions
$P=(1/5)(3/8\pi)^{2/3}h^2/m_n^{8/3}\rho^{5/3}$  (corresponding
to a polytrope of index $n=3/2$) with the relativistic relation $\epsilon=\rho
c^2+(3/2)P$ between the energy density and the rest-mass density. This is the
fermionic counterpart of the partially-relativistic model of BEC stars
(see Appendix \ref{sec_partially}).}
\begin{equation}
\label{mma7}
M_{\rm OV}=0.384 \left (\frac{\hbar c}{G}\right
)^{3/2}\frac{1}{m_n^2}=0.709\, M_{\odot},
\end{equation}
\begin{equation}
\label{mma8}
R_{\rm OV}=8.735  \frac{G M_{OV}}{c^2}=9.15\, {\rm km},
\end{equation}
\begin{equation}
\label{mma9}
(\epsilon_c)_{\rm OV}/c^2=3.44\times 10^{-3} \frac{c^6}{G^3
M_{OV}^2}=4.22\times 10^{15}\, {\rm g}/{\rm cm}^3.
\end{equation}
The mass-radius ratio at the critical point is
\begin{equation}
\label{mma10}
\frac{2GM_{\rm OV}}{R_{\rm OV}c^2}=0.229.
\end{equation}
In this fermionic model, the maximum mass of a neutron star is determined by
fundamental constants and by the mass $m_n$ of the neutrons. As a result, there
is no indetermination and the maximum mass predicted by Oppenheimer and Volkoff
\cite{ov} has a well-specified value $M_{\rm OV}=0.7\, M_{\odot}$.

However, neutron stars with a mass in the range $2-2.4\,  M_{\odot}$, well above
the Oppenheimer-Volkoff limit, have recently been observed
\cite{Lat,Dem,black1,black2,black3}. These observations question the validity of
the fermionic model. Therefore, alternative models of neutron stars should be
constructed. In this respect, Chavanis and Harko \cite{chavharko} have proposed
that, because of their superfluid cores, neutron stars could actually be BEC
stars. Indeed, the neutrons (fermions) could form Cooper pairs  and behave as
bosons of mass $m=2m_n$, where $m_n=0.940\, {\rm GeV}/c^2$ is the mass of the
neutrons. They can then make a BEC through the BCS/BEC crossover  mechanism.
Since the maximum mass of BEC stars $M_{\rm max}=0.307\, \hbar
c^2\sqrt{a_s}/(Gm)^{3/2}$ \cite{chavharko} depends on the scattering length
$a_s$ that is not well-known, it can be larger than the Oppenheimer-Volkoff
limit $M_{OV}=0.376 M_P^3/m_n^2=0.7\, M_{\odot}$ \cite{ov} obtained by assuming
that neutron stars can be modeled as an ideal gas of fermions. By taking a
scattering length of the order of $10-20\, {\rm fm}$ (giving $\kappa\sim
3.16-4.47$), we obtain a maximum mass of the order of $2M_{\odot}$, a central
density of the order $1-3\times 10^{15}\, {\rm g/cm^3}$, and a radius in the
range $10-20\, {\rm km}$ \cite{chavharko}. This could account for the recently
observed neutron stars with masses in the range $2-2.4\, M_{\odot}$ larger than
the Oppenheimer-Volkoff limit. For $M<M_{max}$, there exist stable
equilibrium states of BEC stars with $R>R_{min}$ for which gravitational collapse is prevented by the pressure arising from the scattering length of the bosons. For $M>M_{max}$, nothing prevents the gravitational collapse of the star that becomes a black hole.

It is interesting to come back to the analogies and differences between fermion
and boson stars (in the fully-relativistic model of \cite{chavharko}). In the
ultra-relativistic
limit, they are both described by a polytropic equation of state $P\sim
K'\rho^{4/3}$ corresponding to an index $n=3$ but the polytropic constant is
different. In the case of fermions $K'=(1/4)(3/8\pi)^{1/3}hc/m^{4/3}$ and in the
case of bosons $K'=(\pi/8)^{1/3}\hbar^{2/3}a_s^{1/3}c^{4/3}/m$ (see Appendix
\ref{sec_fully}). In the two cases, the relation between the pressure and the
energy density is $P\sim\epsilon/3$. This is the relation that enters in the TOV
equation. This linear equation of state is responsible for the damped
oscillations of the mass-central density relation and for the snail-like
structure (spiral) of the mass-radius relation. In the non-relativistic limit,
fermion stars are described by a polytropic equation of state $P\sim
K\rho^{5/3}$ of index $n=3/2$. The  polytropic constant is
$K=(1/5)(3/8\pi)^{2/3}h^2/m^{8/3}$. This leads to the mass-radius relation
$MR^3=1.49\times 10^{-3}\, h^6/(G^3 m^8)$. Therefore, there exist configurations
of arbitrarily large radius and arbitrarily small mass (see Fig. \ref{MROV}). By
contrast, in the non-relativistic limit, BEC stars are described by a polytropic
equation of state $P=K\rho^2$ of index $n=1$. The  polytropic constant is
$K=2\pi a_s\hbar^2/m^3$. This fixes the radius of the configuration to the value
$R=\pi(a_s\hbar^2/Gm^3)^{1/2}$. Therefore, there is no configuration of radius
larger than this value (see Fig. \ref{newRMtotal}). This is a difference between
fermion stars and BEC stars. On the other hand, in the case of fermion stars,
the equation of state depends (apart from fundamental constants) only on the
mass $m$ of the fermions. For neutron stars, this is the mass of the neutrons
$m_n$ whose value is perfectly known. Therefore, the maximum mass of neutron
stars modeled as fermion stars has an unambiguous value $0.7\, M_{\odot}$. By
contrast, in the case of BEC stars, the equation of state depends on $m$ and
$a_s$ through the combination  $\kappa^2\propto a_s/m^3$. As a result, the
maximum mass of neutron stars modeled as BEC stars depends on this parameter
$\kappa$  (compare Eqs. (\ref{comp2}) and (\ref{mma7})). Since the value of this
parameter is not well-known, it may be possible to overcome the
Oppenheimer-Volkoff limit.

\subsection{On the maximum mass of baryon stars}
\label{sec_baryon}

Zel'dovich \cite{zeldovich} considered a gas of baryons interacting through a vector meson field and showed that the equation of state of this system is of the form of Eq. (\ref{mm1}) with a polytropic constant
\begin{equation}
\label{baryon1}
K=\frac{g^2 h^2}{2\pi m_m^2 m_b^2 c^2},
\end{equation}
where $g$ is the baryon charge, $m_m$ is the meson mass, and $m_b$ is the baryon
mass. Zel'dovich \cite{zeldovich} introduced this equation of state as an
example to show how the speed of sound could approach the speed of light at very
high pressures and densities (see Appendix \ref{sec_partially}).

The equation of state (\ref{mm1}) has been studied by Tooper \cite{tooper2} in
relation to baryon stars (see his Sec. IX.c). Our treatment is a little more
accurate and provides
the following values for the maximum mass, minimum radius, and maximum density
of baryon stars
\begin{equation}\label{baryon2}
{M_{{\rm max}}}=0.4104\frac{g\hbar c}{m_m m_b G^{3/2}}=3.80\, {M_{\odot}},
\end{equation}
\begin{equation}\label{baryon3}
R_{{\rm min}}=1.914 \frac{g\hbar}{m_m m_b c G^{1/2}}=26.2
\;{\rm km},
\end{equation}
\begin{equation}\label{baryon4}
(\rho_0)_{{\rm max}}=0.318\frac{m_m^2 m_b^2 c^4}{2\pi g^2 \hbar^2}=3.64\times 10^{14}\;{\rm g/cm^3}.
\end{equation}
To make the numerical application, we have taken $g^2/hc\sim 1$, $m_b\sim m_n$, and $m_m=m_b/2$ \cite{tooper2}.

Zel'dovich model is also considered in Sec. 4.2 of \cite{chavbh} in the case
where the equation of state (\ref{mm1}) is approximated by its asymptotic form
$P=\epsilon$ valid at high densities, and the system is enclosed within a
spherical box of radius $R$ to make its mass finite. In this simplified setting,
it is found that the critical mass-radius ratio $2GM/Rc^2$ is equal  to $0.544$
instead of
$0.429$. The agreement is relatively satisfying in view of the crudeness of the
box model. Some analogies between stiff stars and black holes are pointed
out in \cite{chavbh}.

\section{Cosmology of a BEC fluid with a stiff equation of state}
\label{sec_becc}

Harko \cite{harkocosmo} and Chavanis \cite{chavaniscosmo} considered the
possibility that dark matter is made of BECs with a self-interaction and
independently studied the cosmological implications of this
model.\footnote{Harko \cite{harkocosmo} considers repulsive self-interactions
while Chavanis \cite{chavaniscosmo}  considers repulsive and attractive
self-interactions.} If dark matter is made of BECs, it has a non-vanishing
pressure even at $T=0$, unlike the CDM model. This affects the evolution of the
scale factor of the universe. In most applications, the pressure arises from
the self-interaction (the quantum pressure due to the Heisenberg uncertainty
principle is negligible) so we can make the TF approximation.  Harko
\cite{harkocosmo} and Chavanis \cite{chavaniscosmo} consider a polytropic
equation of state of the form of Eq. (\ref{semi1}) and solve the corresponding
Friedmann equations. However, this equation of state is not valid in the
strongly
relativistic regime so the extrapolation of their results to the very early
universe
is not correct.\footnote{Note that the equation of state (\ref{semi1}) is
interesting in its own right in cosmology. Generalized polytropic equations of
state of the form $P=\alpha\epsilon+k\epsilon^{1+1/n}$ have been studied in full
generality in Refs. \cite{cosmopoly1,cosmopoly2,cosmopoly3,quadratic}. For a
negative polytropic pressure ($k<0$), they lead to interesting cosmological
models exhibiting a phase of early inflation and a phase of late accelerated
expansion bridged by a phase of decelerating expansion. However, the
justification of these equations of state may not be connected with BECs as
initially thought.} In this section, we solve the Friedmann equations with the
equation of state (\ref{partially4}). It leads to very different results in the
early universe showing that the precise form of the equation of state of the BEC
is crucial in cosmology.\footnote{It is less crucial in the context of BEC stars
since the precise form of the equation of state only slightly changes the
prefactor of the maximum mass (see Sec. \ref{sec_comp}).} We stress that the
equation of state (\ref{partially4}) is itself not exact so the results of this
section should be considered with caution. However, it is interesting to compare
the effect of different equations of state on the evolution of the universe.
Furthermore, the equation of state (\ref{partially4}) is interesting
because it leads to a cosmological model involving a stiff matter phase. The
same stiff matter phase occurs in the cosmological model of Zel'dovich
\cite{zeldocosmo} where the early universe is assumed to be made of a gas of
cold baryons. The comparison between the different equations of state of BEC
dark matter is made in Sec. \ref{sec_compa}.

\subsection{The Friedmann equations}
\label{sec_friedmann}

We assume that the universe is homogeneous and isotropic, and contains a uniform perfect fluid of energy density $\epsilon(t)$ and isotropic pressure $P(t)$. The radius of curvature of the $3$-dimensional space, or scale factor, is noted $a(t)$ and the curvature of space is noted $k$. The universe is closed if $k>0$, flat if $k=0$, and open if $k<0$. We assume that the universe is flat ($k=0$) in agreement with the observations of the cosmic microwave background (CMB) \cite{bt}. In that case, the Einstein equations can be written as \cite{weinbergbook}:
\begin{equation}
\frac{d\epsilon}{dt}+3\frac{\dot a}{a}(\epsilon+P)=0,
\label{f1}
\end{equation}
\begin{equation}
\frac{\ddot a}{a}=-\frac{4\pi G}{3c^2}\left(\epsilon+3P\right )+\frac{\Lambda}{3},
\label{f2}
\end{equation}
\begin{equation}
H^2=\left (\frac{\dot a}{a}\right )^2=\frac{8\pi G}{3c^2}\epsilon+\frac{\Lambda}{3},
\label{f3}
\end{equation}
where we have introduced the Hubble parameter $H=\dot a/a$ and accounted for
a possible non-zero cosmological constant $\Lambda$. The cosmological constant
is equivalent to a fluid with a constant energy density (dark energy):
\begin{equation}
\epsilon_{\Lambda}=\rho_{\Lambda}c^2=\frac{\Lambda c^2}{8\pi G},
\label{f4}
\end{equation}
and an equation of state $P=-\epsilon$. Eqs. (\ref{f1})-(\ref{f3}) are the
well-known Friedmann equations describing a non-static universe. Among these
three equations, only two are independent. The first equation can be viewed as
an equation of continuity. For a given barotropic equation of state
$P=P(\epsilon)$, it determines the relation between the energy density
$\epsilon$ and the scale factor $a$. Then, the evolution of the scale factor
$a(t)$ is given by Eq. (\ref{f3}).

\subsection{The equation of state of a partially-relativistic BEC fluid}
\label{sec_eos}

We assume that dark matter is made of a fluid at  $T=0$ with an equation of state $P(\rho)$. In that case, the relation between the energy density $\epsilon$ and the mass density $\rho$ is given by the first law of relativistic thermodynamics (see Appendix \ref{sec_gr}):
\begin{equation}
d\epsilon=\frac{P+\epsilon}{\rho}d\rho.
\label{eos1}
\end{equation}
Combining this relation with the continuity equation (\ref{f1}), we get
\begin{equation}
\frac{d\rho}{dt}+3\frac{\dot a}{a}\rho=0.
\label{eos2}
\end{equation}
We note that this equation is exact for a fluid at $T=0$ and that it does not depend on the explicit form of the equation of state $P(\rho)$. It can be integrated into
\begin{equation}
\rho=\rho_0 \left (\frac{a_0}{a}\right )^3,
\label{eos3}
\end{equation}
where $\rho_0$ is the present value of the mass density and $a_0$ is the
present value of the scale factor.

We now assume that dark matter is made of BECs at  $T=0$ described by the
equation of state
\begin{equation}
P=K\rho^2,
\label{eos4}
\end{equation}
where $K$ is given by Eq. (\ref{gen9b}). This polytropic equation of state
of index $n=1$ corresponds to the partially-relativistic model of Appendix
\ref{sec_partially}. The equation of state (\ref{eos4}), with a polytropic
constant $K$ given by Eq. (\ref{baryon1}),  also appears in the cosmological
model of Zel'dovich \cite{zeldocosmo} where the early universe is assumed to be
made of a cold gas of baryons. For the equation of state (\ref{eos4}), Eq.
(\ref{eos1}) can be integrated easily and the relation between the energy
density and the rest-mass density is given by 
\begin{equation}
\epsilon=\rho c^2+K\rho^2.
\label{eos5}
\end{equation}
Combining Eqs. (\ref{eos3}) and (\ref{eos5}), we get
\begin{equation}
\epsilon=\rho_0 c^2\left (\frac{a_0}{a}\right )^3+K\rho_0^2\left (\frac{a_0}{a}\right )^6.
\label{eos6}
\end{equation}
This relation can also be obtained by solving the continuity equation (\ref{f1})
with the equation of state (\ref{partially4}) as detailed in Appendix
\ref{sec_alt}.

In the early universe ($a\rightarrow 0$), we have
\begin{equation}
\epsilon\sim K\rho_0^2\left (\frac{a_0}{a}\right )^6, \qquad \epsilon\sim K\rho^2, \qquad P\sim\epsilon.
\label{eos7}
\end{equation}
These equations describe a stiff fluid ($P=\epsilon$) for which the velocity of sound is equal to the velocity of light.

In the late universe ($a\rightarrow +\infty$), we have
\begin{equation}
\epsilon\sim \rho_0 c^2\left (\frac{a_0}{a}\right )^3, \qquad \epsilon\sim \rho c^2, \qquad P\sim \frac{K}{c^4}\epsilon^2.
\label{eos8}
\end{equation}
These equations describe a classical BEC fluid with a polytropic equation of
state  of index $n=1$ ($P=K\epsilon^2/c^4$). Actually, for very large values of
the scale factor, we recover the results of the CDM model ($P=0$) since
$\epsilon\propto a^{-3}$.

\subsection{The solution of the Friedmann equation for a partially-relativistic BEC fluid}
\label{sec_sol}

Substituting Eq. (\ref{eos6}) in the Friedmann equation (\ref{f3}), we get
\begin{equation}
\left (\frac{\dot a}{a}\right )^2=\frac{8\pi G}{3c^2}\left\lbrack \rho_0 c^2\left (\frac{a_0}{a}\right )^3+K\rho_0^2\left (\frac{a_0}{a}\right )^6+\rho_{\Lambda}c^2\right\rbrack.
\label{s1}
\end{equation}
This first order differential equation determines the temporal evolution of the scale factor $a(t)$. Its formal solution is
\begin{equation}
\int_0^{a(t)/a_0}\frac{dx}{x\sqrt{\frac{\rho_0c^2}{x^3}+\frac{K\rho_0^2}{x^6}+\rho_{\Lambda}c^2}}=\left (\frac{8\pi G}{3c^2}\right )^{1/2}t.
\label{s2}
\end{equation}
This equation has the same form as Eq. (\ref{stiff4}) obtained under the
assumption that the universe is made of three non-interacting fluids
corresponding
to stiff matter, dust matter, and dark energy. Therefore, we can
immediately transpose the results of Appendix \ref{sec_stiff} to the present
context. The universe starts from a primordial singularity and it successively
undergoes a stiff matter phase ($\epsilon\propto a^{-6}$), a dust matter phase
($\epsilon\propto a^{-3}$), and a dark energy phase
($\epsilon\sim\rho_{\Lambda}c^2$).

The evolution of the scale factor is explicitly given by
\begin{eqnarray}
\frac{a}{a_0}=\Biggl \lbrack \left (\frac{\rho_{0}}{\rho_{\Lambda}}+2\sqrt{\frac{K\rho_0^2}{\rho_{\Lambda}c^2}}\right )\sinh^2\left (\sqrt{6\pi G\rho_{\Lambda}} \, t\right )\nonumber\\
+\sqrt{\frac{K\rho_0^2}{\rho_{\Lambda}c^2}}\left (1-e^{-2\sqrt{6\pi G\rho_{\Lambda}}\, t}\right )\Biggr\rbrack^{1/3}.
\label{s3}
\end{eqnarray}
The evolution of the energy density is given by
\begin{eqnarray}
\epsilon&=&\Biggl \lbrack\left
(\frac{1}{2}\sqrt{\frac{\rho_0^2
c^2}{\rho_{\Lambda}}}+\sqrt{K\rho_{0}^2}\right )\sinh \left (2\sqrt{6\pi
G\rho_{\Lambda}}\, t\right )\nonumber\\
&+&\sqrt{K\rho_0^2}e^{-2\sqrt{6\pi G\rho_{\Lambda}}t}\Biggr\rbrack^2/\left
(\frac{a}{a_0}\right )^6 .
\label{s3bb}
\end{eqnarray}
This is a simple generalization of the $\Lambda$CDM model for a BEC universe
assumed to be described by the equation of state (\ref{eos4}). For $t\rightarrow
+\infty$, we recover the de Sitter solution
\begin{eqnarray}
\frac{a}{a_0}\sim \left (\frac{\rho_{0}}{\rho_{\Lambda}}+2\sqrt{\frac{K\rho_0^2}{\rho_{\Lambda}c^2}}\right )^{1/3}\frac{1}{2^{2/3}} e^{\sqrt{\frac{8\pi G\rho_{\Lambda}}{3}}\, t}
\label{s3b}
\end{eqnarray}
with a prefactor affected by the BEC.

In the absence of a cosmological constant ($\rho_{\Lambda}=0$), the solution of
Eq. (\ref{s1}) is
\begin{eqnarray}
\frac{a}{a_0}=\left (6\pi G\rho_0 t^2+2\sqrt{\frac{6\pi GK\rho_0^2}{c^2}}\, t\right )^{1/3},
\label{s4}
\end{eqnarray}
\begin{eqnarray}
\epsilon=\frac{c^2}{6\pi G t^2}\left (\frac{1+\sqrt{\frac{K}{6\pi G}}\frac{1}{c t}}{1+2\sqrt{\frac{K}{6\pi G}}\frac{1}{c t}}\right )^2.
\label{s5}
\end{eqnarray}
This is a simple generalization of the Einstein-de Sitter (EdS) model for a BEC universe assumed to be described by the equation of state (\ref{eos4}).

Returning to the case $\Lambda\ge 0$ and considering the formal limit
$K\rightarrow +\infty$, equivalent to the case of Appendix \ref{sec_stiff} where
we can neglect dust matter in front of stiff matter, we get
\begin{eqnarray}
\frac{a}{a_0}=\left (\frac{K\rho_0^2}{\rho_{\Lambda}c^2}\right )^{1/6}\sinh^{1/3}\left (2 \sqrt{6\pi G\rho_{\Lambda}}\, t\right ),
\label{s6}
\end{eqnarray}
\begin{eqnarray}
\epsilon=\frac{\rho_{\Lambda}c^2}{\sinh^2\left (2\sqrt{6\pi G\rho_{\Lambda}}\, t\right )}.
\label{s7}
\end{eqnarray}
In the absence of cosmological constant ($\Lambda=0$), the foregoing
equations reduce to
\begin{eqnarray}
\frac{a}{a_0}=\left (2\sqrt{\frac{6\pi GK\rho_0^2}{c^2}}\, t\right )^{1/3},\qquad \epsilon=\frac{c^2}{24\pi G t^2}.
\label{s8}
\end{eqnarray}

We also recover well-known models as particular cases of the foregoing
equations. In the absence of BECs ($K=0$), we recover  the $\Lambda$CDM model
\begin{eqnarray}
\frac{a}{a_0}=\left (\frac{\rho_{0}}{\rho_{\Lambda}}\right )^{1/3}\sinh^{2/3}\left (\sqrt{6\pi G\rho_{\Lambda}} \, t\right ),
\label{s9}
\end{eqnarray}
\begin{eqnarray}
\epsilon=\frac{\rho_{\Lambda}c^2}{\sinh^2\left (\sqrt{6\pi G\rho_{\Lambda}}\, t\right )}.
\label{s10}
\end{eqnarray}
In the absence of BECs  and cosmological constant ($K=\Lambda=0$), we recover
the EdS universe
\begin{eqnarray}
\frac{a}{a_0}=\left (6\pi G\rho_0 t^2\right )^{1/3},\qquad \epsilon=\frac{c^2}{6\pi G t^2}.
\label{s11}
\end{eqnarray}

The previous results have been presented in the context of a dark matter fluid
made of BECs with an equation of state given by Eq. (\ref{eos4}). As discussed
in the next section, this model may give wrong results for BECs in the
early universe because it is
based on a {\it classical} equation of state. However, the equation of state
(\ref{eos4}) also appears in the cosmological model of Zel'dovich
\cite{zeldocosmo} where the early universe is assumed to be made of a cold gas
of baryons. This model presents a stiff matter phase that follows the
cosmological singularity (Big Bang). In that context, the equation of state
(\ref{eos4}) is rigorously justified.  Therefore, the analytical solutions that
we have presented in this section are exact in the context of Zel'dovich's model
\cite{zeldocosmo}. Actually, Zel'dovich \cite{zeldocosmo} briefly mentions that
the complete equation of state in his model is of the form
\begin{eqnarray}
P=K\rho^2+K'\rho^{4/3},
\label{compa1}
\end{eqnarray}
where $K$ is given by Eq. (\ref{baryon1}) and the second term accounts for quantum (Fermi) corrections. For the equation of state (\ref{compa1}), we find from Eq. (\ref{gr12}) that the relation between the energy density and the rest-mass density is
\begin{eqnarray}
\epsilon=\rho c^2+K\rho^2+3K'\rho^{4/3}.
\label{compa2}
\end{eqnarray}
Substituting Eq. (\ref{eos3}) in Eq. (\ref{compa2}), we obtain
\begin{equation}
\epsilon=\rho_0 c^2\left (\frac{a_0}{a}\right )^3+K\rho_0^2\left (\frac{a_0}{a}\right )^6+3K'\rho_0^{4/3}\left (\frac{a_0}{a}\right )^4.
\label{compa3}
\end{equation}
When combined with the Friedmann equation  (\ref{f3}), we obtain a model of universe exhibiting a stiff matter phase ($\epsilon\propto a^{-6}$), a radiation phase ($\epsilon\propto a^{-4}$), a dust matter phase ($\epsilon\propto a^{-3}$), and a dark energy phase ($\epsilon\sim\rho_{\Lambda}c^2$) as discussed in Appendix \ref{sec_stiff}.

{\it Remark:} In this paper, we have considered the case of a
repulsive self-interaction
($K\ge 0$). The case of an attractive self-interaction ($K<0$) is treated in
\cite{stiff}. In that case, the primordial universe is non-singular. We
have also assumed that the cosmological constant is positive in agreement with
the observations. The case of a negative cosmological constant (anti-de Sitter),
leading to an
oscillatory universe, is considered in \cite{stiff}.

\subsection{Comparison between the different models of BEC cosmology}
\label{sec_compa}

We now compare the different models of BEC cosmology depending on the considered equation of state.

Harko \cite{harkocosmo} and Chavanis \cite{chavaniscosmo}  assumed that the BEC
dark matter is described by an equation of state of the form (\ref{semi1})  and
solved
the corresponding Friedmann equations. This corresponds to the semi-relativistic
model of Appendix \ref{sec_semi}. For a repulsive self-interaction ($a_s>0$)
they found that the universe starts with a new form of singularity in which the
energy density is infinite while the scale factor is finite. At sufficiently
late times, the universe returns the usual $\Lambda$CDM model in which the
universe experiences an EdS
phase ($a\propto t^{2/3}$, $\epsilon\propto t^{-2}$) followed by a de Sitter
phase ($a\propto e^{\sqrt{\Lambda/3}t}$, $\epsilon=\rho_{\Lambda} c^2$).
However, this model differs from the $\Lambda$CDM model in the intermediate
phase because of the contribution of the BEC. In particular, it is found that
the scale factor increases more rapidly when dark matter is made of BEC instead
of pressureless matter \cite{harkocosmo,chavaniscosmo}.

In this paper, we have assumed that the BEC dark matter is described by the
equation
of state (\ref{partially1}). This corresponds to the partially-relativistic
model of Appendix \ref{sec_partially}. Solving the corresponding Friedmann
equation, we have found that the early universe behaves as a stiff fluid ($P\sim
\epsilon$). The scale factor increases as $a\propto t^{1/3}$ while the energy
density decreases as $\epsilon\propto t^{-2}$. The universe starts from a
singularity at $t=0$ in which the energy density is infinite while the scale
factor vanishes. At later  times, the universe behaves as a nonrelativistic BEC
with an equation of state $P\sim K\epsilon^2/c^4$. 

We stress, however, that the previous models may be incorrect in the very early
universe because they use an approximate relativistic equation of state. A
better model of BEC dark matter should be based on the equation of state
(\ref{fully1}).  This corresponds to the fully-relativistic model of Appendix
\ref{sec_fully}. In that case, the early universe has an equation of state
$P\sim\epsilon/3$ similar to the equation of state of radiation. At later 
times, the universe behaves as a nonrelativistic BEC with an
equation of state $P\sim K\epsilon^2/c^4$. 

{\it Remark:} Although the evolution of the early universe is very sensitive to
the equation of state of the BEC, it should be recalled that BECs do form only
when the temperature has sufficiently decreased. Therefore, we should be careful
when extrapolating the solutions to the past. If we view BEC dark matter as a
small correction to pressureless matter ($\Lambda$CDM model),
all the equations of state of Appendix \ref{sec_beceos} reduce to $P\sim
K\epsilon^2/c^4$ and give equivalent results for sufficiently late times. In
this sense, the cosmological models of Harko \cite{harkocosmo} and Chavanis
\cite{chavaniscosmo} are justified for sufficiently late times after the
primordial singularity and after the appearance of BECs. 

\section{Conclusion}
\label{sec_con}

In this paper, we have compared different models of relativistic
self-gravitating BECs.

Concerning general relativistic BEC stars, we have shown that
the partially-relativistic model of Appendix \ref{sec_partially} (leading to a
stiff equation of state) gives a maximum mass that is smaller than the
semi-relativistic model of Appendix \ref{sec_semi} but larger than the
fully-relativistic model of Appendix \ref{sec_fully}. However, the difference is
relatively small (the main indetermination of the maximum mass being the value
of the scattering length of the particles) so that the three models provide a
fair description of general relativistic BEC stars. Of course, the fully
relativistic treatment is the most relevant one on a physical
point of view. However, the observed mass-radius ratio of neutron stars seems to
be closer to
the value obtained from the partially-relativistic model or from the value
obtained from the semi-relativistic model than to the value obtained from
the fully-relativistic model. Therefore, the equations of state
(\ref{partially4}) and
(\ref{semi1})  may be useful to model neutron stars,
independently of the
BEC model. In this respect, we can note that they are special cases of the two
polytropic models developed by Tooper \cite{tooper1,tooper2} for an
index $n=1$. However, more observations may be necessary to determine a precise
value of the mass-radius ratio of neutron stars and ascertain these conclusions.

Concerning the evolution of a universe made of BEC dark matter, the precise form
of the equation of state is crucial in the very early universe. If the dark
matter is
described by the equation of state (\ref{semi1}), corresponding to the
semi-relativistic model, the universe starts from a primordial singularity in
which the scale factor is finite and the density is infinite. On the other hand,
if the dark matter is described by the equation of state (\ref{partially4})
corresponding to the partially relativistic model, the early universe undergoes
a stiff matter era in which the scale factor increases as $a \propto t^{1/3}$ 
and the energy density decreases as $\epsilon\propto a^{-6}$. Finally, if the
dark matter is described by the equation of state (\ref{fully1}) corresponding
to the fully relativistic model, the early universe undergoes a radiation era
in which the scale factor increases as  $a \propto t^{1/2}$ and the energy
density decreases as $\epsilon\propto a^{-4}$. In principle, only the
fully-relativistic model is relevant in the very early universe. However, at
later
times, all the models give equivalent results.

The stiff equation of state (\ref{partially4}) also describes a baryon star or
the primordial evolution of a universe filled with a cold gas of baryons as
proposed by Zel'dovich \cite{zeldocosmo,zeldovich}. Therefore, the  results that
we obtained with this equation of state can have application in the context of
cold baryons (Zel'dovich model), independently of the BEC model.


\appendix

\section{The equation of state of a relativistic BEC}
\label{sec_beceos}

\subsection{General results}
\label{sec_gr}

The local form of the first law of thermodynamics can be expressed as
\begin{equation}
d\left (\frac{\epsilon}{\rho}\right )=-P d\left (\frac{1}{\rho}\right )+T d\left (\frac{s}{\rho}\right ),
\label{gr1}
\end{equation}
where $\rho=n m$ is the mass density, $n$ is the number density, and $s$ is the entropy density in the rest frame. For a system at $T=0$, the first law of thermodynamics reduces to
\begin{equation}
d\epsilon=\frac{P+\epsilon}{\rho}d\rho.
\label{gr2}
\end{equation}
For a given equation of state, Eq. (\ref{gr2}) can be integrated to obtain the
relation between the energy density $\epsilon$ and the rest-mass density $\rho$.

If the equation of state is prescribed under the form  $P=P(\epsilon)$, Eq. (\ref{gr2}) can be immediately integrated into
\begin{equation}
\ln\rho=\int \frac{d\epsilon}{P(\epsilon)+\epsilon}.
\label{gr3}
\end{equation}
If, as an example, we consider the ``gamma law'' equation of state \cite{bondi,htww}:
\begin{equation}
P=(\gamma-1)\epsilon,
\label{gr4}
\end{equation}
we get
\begin{equation}
P=K\rho^{\gamma},\qquad \epsilon=\frac{K}{\gamma-1}\rho^{\gamma},
\label{gr5}
\end{equation}
where $K$ is a constant of integration.

We now assume that the equation of state is prescribed under the form $P=P(\rho)$. In that case, Eq. (\ref{gr2}) reduces to the first order linear differential equation
\begin{equation}
\frac{d\epsilon}{d\rho}-\frac{1}{\rho}\epsilon=\frac{P(\rho)}{\rho}.
\label{gr6}
\end{equation}
Using the method of the variation of the constant, we obtain
\begin{equation}
\epsilon=A\rho c^2+\rho\int^{\rho}\frac{P(\rho')}{{\rho'}^2}\, d\rho',
\label{gr7}
\end{equation}
where $A$ is a constant of integration.

As an example, we consider the polytropic equation of state \cite{chandra}:
\begin{equation}
P=K\rho^{\gamma}, \qquad \gamma=1+\frac{1}{n}.
\label{gr8}
\end{equation}
For $\gamma=1$, we get
\begin{equation}
\epsilon=A\rho c^2+K\rho\ln\rho.
\label{gr9}
\end{equation}
For $\gamma\neq 1$, we obtain
\begin{equation}
\epsilon=A\rho c^2+\frac{K}{\gamma-1}\rho^{\gamma}=A\rho c^2+nP.
\label{gr10}
\end{equation}
Taking $A=0$, we recover Eqs. (\ref{gr4})-(\ref{gr5}). We now assume
$0<n<+\infty$ (i.e. $\gamma>1$). In that case, we determine the constant $A$ by
requiring that $\epsilon\sim\rho c^2$ when $\rho\rightarrow 0$. This gives
$A=1$. As a result, Eq. (\ref{gr10}) takes the form
\begin{equation}
\epsilon=\rho c^2+\frac{K}{\gamma-1}\rho^{\gamma}=\rho c^2+nP.
\label{gr11}
\end{equation}
For $\rho\rightarrow 0$ (non-relativistic limit), we get
\begin{equation}
\epsilon\sim \rho c^2,\qquad P\sim K(\epsilon/c^2)^{\gamma}.
\label{gr11b}
\end{equation}
For  $\rho\rightarrow +\infty$ (ultra-relativistic limit), we get
\begin{equation}
\epsilon\sim n K \rho^{\gamma},\qquad P\sim \epsilon/n\sim (\gamma-1)\epsilon.
\label{gr11c}
\end{equation}

For a general equation of state $P(\rho)$ such that $P\sim \rho^{\gamma}$ with $\gamma>1$ when $\rho\rightarrow 0$, we  determine the constant $A$ in Eq. (\ref{gr7}) by requiring that $\epsilon\sim\rho c^2$ when $\rho\rightarrow 0$. This gives
\begin{equation}
\epsilon=\rho c^2+\rho\int_0^{\rho}\frac{P(\rho')}{{\rho'}^2}\, d\rho'=\rho c^2+u(\rho).
\label{gr12}
\end{equation}
We note that $\rho c^2$ is the rest mass energy density and $u(\rho)$ may be
interpreted as an internal energy \cite{aaantonov}.

\subsection{Fully-relativistic model}
\label{sec_fully}

We consider the equation of state
\begin{equation}
\label{fully1}
P=\frac{c^4}{36K}\left (\sqrt{1+\frac{12K}{c^4}\epsilon}-1\right )^2,
\end{equation}
where $K$ is given by Eq.~(\ref{gen9b}). This equation of state can be derived
from the relativistic Klein-Gordon equation of a self-interacting scalar field
in the strong coupling limit \cite{colpi}. It also applies to a relativistic
self-interacting BEC at $T=0$ in the TF approximation \cite{chavharko}. It
provides a fully-relativistic BEC model. For $\epsilon \rightarrow 0$
(non-relativistic limit), we recover the polytropic equation of state
$P=K(\epsilon/c^2)^2$ of a classical BEC. For $\epsilon \rightarrow +\infty$
(ultra-relativistic limit), we obtain a linear equation of state  $P=\epsilon/3$
similar to the one describing the core of neutron stars modeled by the ideal
Fermi gas (see Sec. \ref{sec_mma}) \cite{ov,misner,mt,chavgen,chavbh}.

For the equation of state (\ref{fully1}), Eq. (\ref{gr3}) becomes
\begin{equation}
\frac{1}{3}\ln\rho=\int^{12K\epsilon/c^4} \frac{dx}{\left (\sqrt{1+x}-1\right )^2+3x}.
\label{fully2}
\end{equation}
Using the identity
\begin{eqnarray}
\int \frac{dx}{(\sqrt{1+x}-1)^2+3x}=\frac{1}{3}\ln(\sqrt{1+x}-1)\nonumber\\
+\frac{1}{6}\ln(1+2\sqrt{1+x}),
\label{fully3}
\end{eqnarray}
and requiring that  $\epsilon\sim\rho c^2$ for $\rho\rightarrow 0$, we obtain the following relation between the rest-mass density and the energy density
\begin{equation}
\rho=\frac{c^2}{6\sqrt{3}K}\left (\sqrt{1+\frac{12K}{c^4}\epsilon}-1\right )\left\lbrack 1+2\sqrt{1+\frac{12K}{c^4}\epsilon}\right\rbrack^{1/2}.
\label{fully4}
\end{equation}

For $\rho\rightarrow 0$ (non-relativistic limit), we get
\begin{eqnarray}
\epsilon\sim \rho c^2,\qquad P\sim \frac{K}{c^4}\epsilon^2,\qquad P\sim K\rho^2,
\label{fully5}
\end{eqnarray}
corresponding to a polytrope $n=1$. This returns the equation of state (\ref{gen8}) of a classical BEC. For $\rho\rightarrow +\infty$ (ultra-relativistic limit), we get
\begin{eqnarray}
\epsilon\sim \frac{3 c^{4/3}K^{1/3}}{2^{4/3}}\rho^{4/3},\quad P\sim \frac{1}{3}\epsilon,\quad P\sim \frac{c^{4/3}K^{1/3}}{2^{4/3}}\rho^{4/3},\nonumber\\
\label{fully6}
\end{eqnarray}
corresponding to a polytrope $n=3$. This is similar to the equation of state of
an ultra-relativistic 
Fermi gas at $T=0$ (core of neutron star) but the polytropic constant is
different (see Sec. \ref{sec_mma}).

For the equation of state (\ref{fully1}), the velocity of sound is given by
\begin{equation}
\label{fully7}
\frac{c_s^2}{c^2}=P'(\epsilon)=\frac{1}{3}\left (1-\frac{1}{\sqrt{1+12K\epsilon/c^4}}\right).
\end{equation}
We always have $c_s< c$. For $\epsilon\rightarrow +\infty$, $c_s\rightarrow
c/\sqrt{3}$.

\subsection{Partially-relativistic model}
\label{sec_partially}

We consider the equation of state
\begin{equation}
P=K\rho^{2},
\label{partially1}
\end{equation}
where $K$ is given by Eq.~(\ref{gen9b}). This equation of state can be derived
from the classical GP equation. It describes a non-relativistic self-interacting
BEC at $T=0$ in the TF approximation. We assume that this relation remains valid
in the relativistic regime. This is not exact but it provides a
partially-relativistic BEC model.

Since the equation of state (\ref{partially1}) corresponds to a polytrope $n=1$, Eq. (\ref{gr11}) reduces to
\begin{equation}
\epsilon=\rho c^2+P=\rho c^2+K\rho^2.
\label{partially2}
\end{equation}
This equation can be reversed to give
\begin{equation}
\rho=\frac{c^2}{2K}\left (\sqrt{1+\frac{4K\epsilon}{c^4}}-1\right ).
\label{partially3}
\end{equation}
Combining Eqs. (\ref{partially1}) and (\ref{partially3}), we obtain the relation between the pressure and the energy density
\begin{equation}
P=\frac{c^4}{4K}\left (\sqrt{1+\frac{4K\epsilon}{c^4}}-1\right )^2.
\label{partially4}
\end{equation}
This equation of state has a form similar to Eq. (\ref{fully1}) but the
coefficients are different (see Appendix \ref{sec_diff}). We note that Eq.
(\ref{partially1}) with Eq. (\ref{partially2}) is a particular case of the class
of equations of state studied by Tooper \cite{tooper2} in general relativity.
For $\epsilon \rightarrow 0$ (non-relativistic limit), we recover the polytropic equation of state
$P=K(\epsilon/c^2)^2$ of a classical BEC. For $\epsilon \rightarrow +\infty$
(ultra-relativistic limit), we obtain a linear equation of state $P=\epsilon$.
This is a stiff equation of state in which the velocity of sound is equal to the
velocity of light ($c_s=c$). This type of equations of state has been introduced
by Zel'dovich \cite{zeldovich}  in the context of baryon stars in which the
baryons interact through a vector meson field (see Sec. \ref{sec_baryon}).

For $\rho\rightarrow 0$ (non-relativistic limit), we get
\begin{eqnarray}
\epsilon\sim \rho c^2,\qquad P\sim \frac{K}{c^4}\epsilon^2,\qquad P= K\rho^2.
\label{partially5}
\end{eqnarray}
For $\rho\rightarrow +\infty$ (ultra-relativistic limit), we get
\begin{eqnarray}
\epsilon\sim K\rho^{2},\qquad P\sim \epsilon,\qquad P= K\rho^{2}.
\label{partially6}
\end{eqnarray}

For the equation of state (\ref{fully1}), the velocity of sound is given by
\begin{equation}
\label{partially7}
\frac{c_s^2}{c^2}=P'(\epsilon)=1-\frac{1}{\sqrt{1+4K\epsilon/c^4}}.
\end{equation}
We always have $c_s\le c$. For $\epsilon\rightarrow +\infty$,  $c_s\rightarrow
c$.

\subsection{Semi-relativistic model}
\label{sec_semi}

We consider the equation of state
\begin{equation}
\label{semi1}
P=\frac{K}{c^4}\epsilon^2,
\end{equation}
where $K$ is given by Eq.~(\ref{gen9b}). This equation of state was studied as a
preliminary model by Chavanis and Harko \cite{chavharko} before treating the
fully-relativistic model corresponding to the equation of state (\ref{fully1}).
The equation of state (\ref{semi1}) is obtained from Eq. (\ref{gen8}), derived
from the classical GP equation, by replacing the rest mass density by the energy
density (i.e. by making the approximation $\epsilon\sim \rho c^2$), and
by assuming that the resulting equation remains valid in the relativistic
regime. This is not exact but it provides a semi-relativistic BEC model. We note
that Eq. (\ref{semi1}) is a particular case of the class of polytropic equations
of state studied by Tooper \cite{tooper1} in general relativity.

For the equation of state (\ref{semi1}), Eq. (\ref{gr3}) becomes
\begin{equation}
\ln\rho=\int \frac{d\epsilon}{\left (\frac{K\epsilon}{c^4}+1\right )\epsilon}.
\label{semi2}
\end{equation}
Performing the integral and requiring that $\epsilon\sim\rho c^2$ for $\rho\rightarrow 0$, we obtain the following relation between the mass density and the energy density
\begin{equation}
\rho=\frac{\epsilon/c^2}{\frac{K\epsilon}{c^4}+1}.
\label{semi3}
\end{equation}
This equation can be reversed to give
\begin{equation}
\epsilon=\frac{\rho c^2}{1-\frac{K\rho}{c^2}}.
\label{semi4}
\end{equation}
Combining Eqs. (\ref{semi1}) and (\ref{semi4}) we obtain
\begin{equation}
P=\frac{K\rho^2}{(1-K\rho/c^2)^2}.
\label{semi5}
\end{equation}
We note that the pressure diverges when $\rho=c^2/K$. Therefore, there is a maximum density
\begin{equation}
\rho_{max}=\frac{c^2}{K}=\frac{m^3 c^2}{2\pi a_s \hbar^2}.
\label{semi6}
\end{equation}
For $\rho\rightarrow 0$ (non-relativistic regime), we get
\begin{eqnarray}
\epsilon\sim \rho c^2,\qquad P\sim \frac{K}{c^4}\epsilon^2,\qquad P\sim K\rho^2.
\label{semi7}
\end{eqnarray}
For $\rho\rightarrow c^2/K$, we get
\begin{eqnarray}
\epsilon\sim \frac{c^4/K}{1-\frac{K\rho}{c^2}},\qquad P\sim \frac{K}{c^4}\epsilon^2,\qquad P\sim \frac{c^4/K}{(1-K\rho/c^2)^2}.\nonumber\\
\label{semi8}
\end{eqnarray}

For the equation of state (\ref{semi1}), the velocity of sound is given by
\begin{equation}
\label{semi9}
c_s^2=\frac{2K}{c^2}\epsilon.
\end{equation}
The velocity of sound can be mathematically larger than the velocity of light
but such configurations are dynamically unstable \cite{chavharko}.

{\it Remark:} For a general polytropic equation of state of the form
$P=K(\epsilon/c^2)^{\gamma}$ we get 
\begin{equation}
\label{semi10}
\rho=\frac{\epsilon/c^2}{\left\lbrack
1+\frac{K}{c^{2\gamma}}\epsilon^{\gamma-1}\right
\rbrack^{\frac{1}{\gamma-1}}},\quad \epsilon=\frac{\rho c^2}{\left
(1-\frac{K}{c^2}\rho^{\gamma-1}\right )^{\frac{1}{\gamma-1}}},
\end{equation}
and
\begin{equation}
\label{semi11}
P=\frac{K\rho^{\gamma}}{\left (1-\frac{K}{c^2}\rho^{\gamma-1}\right
)^{\frac{\gamma}{\gamma-1}}}.
\end{equation}

\section{The Newtonian value of the maximum mass of a relativistic BEC star}
\label{sec_newcolp}

It is interesting to compare the value of the maximum mass of a relativistic BEC
star obtained by using general relativity (see \cite{chavharko}) with the one
obtained by using Newtonian gravity. We consider the fully relativistic equation
of state of Appendix \ref{sec_fully}. In general relativity, we have to
substitute in the TOV equations (\ref{tov1}) and (\ref{tov2}) the equation of
state (\ref{fully1}) relating the pressure $P$ to the energy density $\epsilon$.
This leads to the maximum mass of Eq. (\ref{comp2}) with $A_2=0.307$
\cite{chavharko}. In Newtonian gravity, we have to substitute in the equation of
hydrostatic equilibrium (\ref{tf1}) the equation of state defined by Eqs.
(\ref{fully1}) and (\ref{fully4}) relating the pressure $P$ to the mass density
$\rho$. In order to determine the maximum mass of a relativistic  BEC star in
the Newtonian framework, it is sufficient to consider the limiting form of this
equation of state in the ultra-relativistic limit $\rho\rightarrow
+\infty$.\footnote{We use a treatment similar to the one performed by
Chandrasekar \cite{chandrasimple} in the case of relativistic white dwarf stars
treated with Newtonian gravity.} Eq. (\ref{fully6}) can be rewritten as
\begin{eqnarray}
P=\frac{\pi^{1/3}c^{4/3}\hbar^{2/3}a_s^{1/3}}{2m}\rho^{4/3}.
\label{newcolp1}
\end{eqnarray}
This is the equation of state of a polytrope of index $n=3$ \cite{chandra}. 
Combining the condition of hydrostatic equilibrium (\ref{tf1}) with the Poisson
equation (\ref{gen6}), we obtain
\begin{eqnarray}
\nabla\cdot \left (\frac{\nabla P}{\rho}\right )=-4\pi G\rho.
\label{newcolp2}
\end{eqnarray}
Substituting the equation of state (\ref{newcolp1}) in Eq. (\ref{newcolp2}), we get
\begin{eqnarray}
\frac{c^{4/3}\hbar^{2/3}a_s^{1/3}}{2\pi^{2/3}Gm}\Delta\rho^{1/3}=-\rho.
\label{newcolp3}
\end{eqnarray}
Defining 
\begin{eqnarray}
\rho=\rho_0\theta^3,\qquad \xi=\left (\frac{2\pi^{2/3}Gm\rho_0^{2/3}}{c^{4/3}\hbar^{2/3}a_s^{1/3}}\right )^{1/2}r,
\label{newcolp3b}
\end{eqnarray}
where $\rho_0$ is the central density, we obtain the Lane-Emden equation
\begin{equation}
\label{newcolp4}
\frac{1}{\xi ^{2}}\frac{d}{d\xi }\left (\xi ^{2}\frac{d\theta }{d\xi }\right )=-\theta^3,
\end{equation}
\begin{equation}
\label{newcolp5}
\theta(0)=1,\qquad \theta'(0)=0,
\end{equation}
for a polytrope of index $n=3$ \cite{chandra}. This equation has to be solved
numerically. The function $\theta(\xi)$ vanishes at $\xi_1=6.89685$. At that
point $\omega_3\equiv -\xi_1^2\theta'(\xi_1)=2.01824$. The mass of the star
$M=\int_{0}^{R} \rho 4\pi r^2\, dr$ is given by
\begin{equation}
\label{newcolp6}
M=\sqrt{2}\omega_3\, \frac{\hbar c^2\sqrt{a_s}}{(Gm)^{3/2}},
\end{equation}
with $\sqrt{2}\omega_3=2.854$. We note that the radius $R$ of a star described
by the equation of state (\ref{newcolp1}) can take arbitrary values. The
relation between the radius and the central density of the star is
\begin{equation}
\label{newcolp7}
R=\frac{\xi_1}{\sqrt{2}\pi^{1/3}}\, \frac{c^{2/3}\hbar^{1/3}a_s^{1/6}}{m^{1/2}G^{1/2}\rho_0^{1/3}},
\end{equation}
with $\xi_1/\sqrt{2}\pi^{1/3}=3.33$. We have the relation 
\begin{equation}
\label{newcolp8}
\rho_0 R^3=\frac{\xi_1^3}{4\pi\omega_3}\, M,
\end{equation}
with ${\xi_1^3}/{4\pi\omega_3}=12.9$. The mean density of the star
$\overline{\rho}=3M/4\pi R^{3}$ is related to the central density by the
relation $\overline{\rho}=3\omega_3\rho _{0}/\xi_1^3=1.85\times 10^{-2}\rho_0$.

It is interesting to contrast the calculations of this Appendix to those of Sec.
\ref{sec_tf}. For a polytrope of index $n=1$, the radius of the star is fixed
but its mass is unspecified. Inversely, for a polytrope of index $n=3$, the
mass of the star is fixed but its radius is unspecified. For other values of the
polytropic index, the radius of the star is a function of its mass. The general
mass-radius relation of a polytropic star with an equation of state
$P=K\rho^{1+1/n}$ is \cite{chandra}:
\begin{equation}
\label{newcolp9}
M^{(n-1)/n}R^{(3-n)/n}=\frac{K(n+1)}{G(4\pi)^{1/n}}\omega_n^{(n-1)/n},
\end{equation}
where $\omega_n=-\xi_1^{(n+1)/(n-1)}\theta'(\xi_1)$. For $n=1$ and $n=3$, we
recover the results of Sec.  \ref{sec_tf} and of the present Appendix.

When the full equation of state $P(\rho)$ defined by Eqs. (\ref{fully1}) and
(\ref{fully4}) is considered, we find that the limiting configuration
corresponding to the ultra-relativistic limit determined by the equation of
state (\ref{newcolp1}) has a radius $R=0$ and an infinite density
$\rho_0\rightarrow +\infty$ \cite{prep}. This configuration corresponds to a
Dirac peak of mass $M$. Therefore, the mass defined by Eq. (\ref{newcolp6})
corresponds to the maximum mass of a relativistic BEC star in Newtonian gravity.
It has the correct scaling of Eq. (\ref{grz1}) but the prefactor $2.854$ is very
different from the exact prefactor $0.307$ obtained in general relativity. This
shows that general relativity is crucial to determine the maximum mass of
relativistic BEC stars. The same conclusion is
reached for neutron stars considered as fermion stars. The general relativistic
approach of Oppenheimer and Volkoff \cite{ov} leads
to a maximum mass equal to $M_{max}=0.376\, (\hbar
c/G)^{3/2}m_n^{-2}=0.7\, M_{\odot}$ while the Newtonian treatment of
Chandrasekhar leads to a maximum
mass equal to $M'_{max}=3.10\, (\hbar c/G)^{3/2}m_n^{-2}=5.76\, M_{\odot}$
\cite{chandrasimple,chandracomplet}.

\section{Generalized equation of state and alternative form of the differential
equation (\ref{mm6})-(\ref{mm7})}
\label{sec_diff}

Let us consider the equation of state
\begin{equation}
\label{diff1}
P=\frac{q^2c^4}{4K}\left (\sqrt{1+\frac{4K}{qc^4}\epsilon}-1\right )^2.
\end{equation}
When $\epsilon\rightarrow 0$ (non-relativistic regime) we recover the quadratic
equation of state $P\sim K\epsilon^2/c^4$ of a classical BEC. When
$\epsilon\rightarrow +\infty$ (ultra-relativistic regime) we obtain a linear
equation of state $P\sim q\epsilon$. The equation of state (\ref{diff1})
generalizes the equations of state (\ref{fully1}) and (\ref{partially4})
corresponding to $q=1/3$ and $q=1$ respectively.

The relation between the rest-mass density and the energy density is given by Eq. (\ref{gr3}) which can be rewritten as
\begin{equation}
q\ln\rho=\int^{4 K\epsilon/q c^4} \frac{dx}{\left (\sqrt{1+x}-1\right )^2+\frac{x}{q}}.
\label{diff2}
\end{equation}
The integral can be calculated analytically. Setting $y=\sqrt{x+1}$, we obtain
\begin{equation}
\int \frac{dx}{\left (\sqrt{1+x}-1\right )^2+\frac{x}{q}}=\frac{A}{B},
\label{diff2a}
\end{equation}
with
\begin{eqnarray}
A=q(y-1)\lbrack 1+y+q(y-1)\rbrack \nonumber\\
\times\lbrace (1+q)\ln(y-1)-(q-1)\ln\lbrack 1+y+q(y-1)\rbrack\rbrace
\label{diff2b}
\end{eqnarray}
and
\begin{equation}
B=(1+q)\lbrack x+q(2+x-2y)\rbrack.
\label{diff2c}
\end{equation}
For $q=1/3$, we recover Eq. (\ref{fully3}), and for $q=1$ the expression in the
right hand side of Eq. (\ref{diff2}) reduces to $\ln(y-1)$ returning the
equations of Appendix \ref{sec_partially} obtained the other way round.

In Sec. \ref{sec_mm}, we have written the TOV equations associated to
the equation of state (\ref{mm1}), equivalent to Eq. (\ref{mm2}), by introducing
the variable $\theta$ defined with the rest-mass density $\rho$. Alternatively,
we can introduce a variable $\Theta$ defined with the energy density $\epsilon$
 as in \cite{chavharko} (be careful to the change of notations). If we set
\begin{equation}\label{mm4w}
\epsilon=\epsilon_0 \Theta,\quad r=\frac{\xi}{A}, \quad \sigma=\frac{K\epsilon_0}{c^4},
\end{equation}
\begin{equation}\label{mm5w}
M(r)=\frac{4\pi\epsilon_0}{A^3c^2}v(\xi), \quad A=\left (\frac{2\pi G}{K}\right )^{1/2},
\end{equation}
and substitute the equation of state (\ref{diff1}) in the TOV equations
(\ref{tov1}) and (\ref{tov2}), using
\begin{equation}
\label{mm23b}
P'(\epsilon)=q\left\lbrack 1-\frac{1}{\sqrt{1+4K\epsilon/qc^4}}\right\rbrack,
\end{equation}
we obtain
\begin{widetext}
\begin{equation}\label{mm24}
\frac{d\Theta}{d\xi}=-\frac{\frac{2}{q}\left\lbrack \frac{q^2}{4}(\sqrt{1+4\sigma\Theta/q}-1)^2+\sigma\Theta\right\rbrack\left\lbrack v+\frac{q^2\xi^3}{4\sigma}(\sqrt{1+4\sigma\Theta/q}-1)^2\right\rbrack}{\xi^2(1-4\sigma v/\xi)(1-1/\sqrt{1+4\sigma\Theta/q})},
\end{equation}
\begin{equation}\label{mm25}
\frac{dv}{d\xi}=\Theta \xi^2.
\end{equation}
\end{widetext}
For $q=1/3$ these equations reduce to Eqs. (87) and (88) of \cite{chavharko}.

\section{Alternative derivation of Eq. (\ref{eos6})}
\label{sec_alt}

In this Appendix, we check that Eq. (\ref{eos6}) can be obtained directly from the equation of continuity (\ref{f1}) with the equation of state (\ref{partially4}).

Substituting Eq. (\ref{partially4}) in Eq. (\ref{f1}), and simplifying some terms, we get
\begin{equation}
\label{alt1}
\frac{2K}{c^4}\frac{d\epsilon}{da}+\frac{3}{a}\left (\frac{4K\epsilon}{c^4}+1-\sqrt{1+\frac{4K\epsilon}{c^4}}\right )=0.
\end{equation}
With the change of variables $x=(1+4K\epsilon/c^4)^{1/2}$, we obtain
\begin{equation}
\label{alt2}
\frac{dx}{da}+\frac{3}{a}(x-1)=0.
\end{equation}
This equation can be integrated into
\begin{equation}
\label{alt3}
x=1+\frac{A}{a^3},
\end{equation}
where $A$ is a constant. Returning to original variables, we obtain
\begin{equation}
\label{alt4}
\epsilon=\frac{c^4}{4K}\left (\frac{2A}{a^3}+\frac{A^2}{a^6}\right ),
\end{equation}
which can be written as Eq. (\ref{eos6}).

\section{A universe with a stiff matter era}
\label{sec_stiff}

In this Appendix, we assume that the universe is made of one or several
fluids each of them described by a linear equation of state $P=\alpha\epsilon$.
The equation of continuity (\ref{f1}) implies that the energy density is related
to the scale factor by $\epsilon=\epsilon_0 (a_0/a)^{3(1+\alpha)}$, where the
subscript $0$ denotes present-day values of the quantities. A linear equation of
state can describe dust matter ($\alpha=0$, $\epsilon_m\propto a^{-3}$),
radiation ($\alpha=1/3$, $\epsilon_{rad}\propto a^{-4}$), stiff matter
($\alpha=1$, $\epsilon_s\propto a^{-6}$), vacuum energy ($\alpha=-1$,
$\epsilon=\epsilon_P$), and dark energy ($\alpha=-1$,
$\epsilon=\epsilon_{\Lambda}$).

\begin{figure}[!ht]
\includegraphics[width=0.98\linewidth]{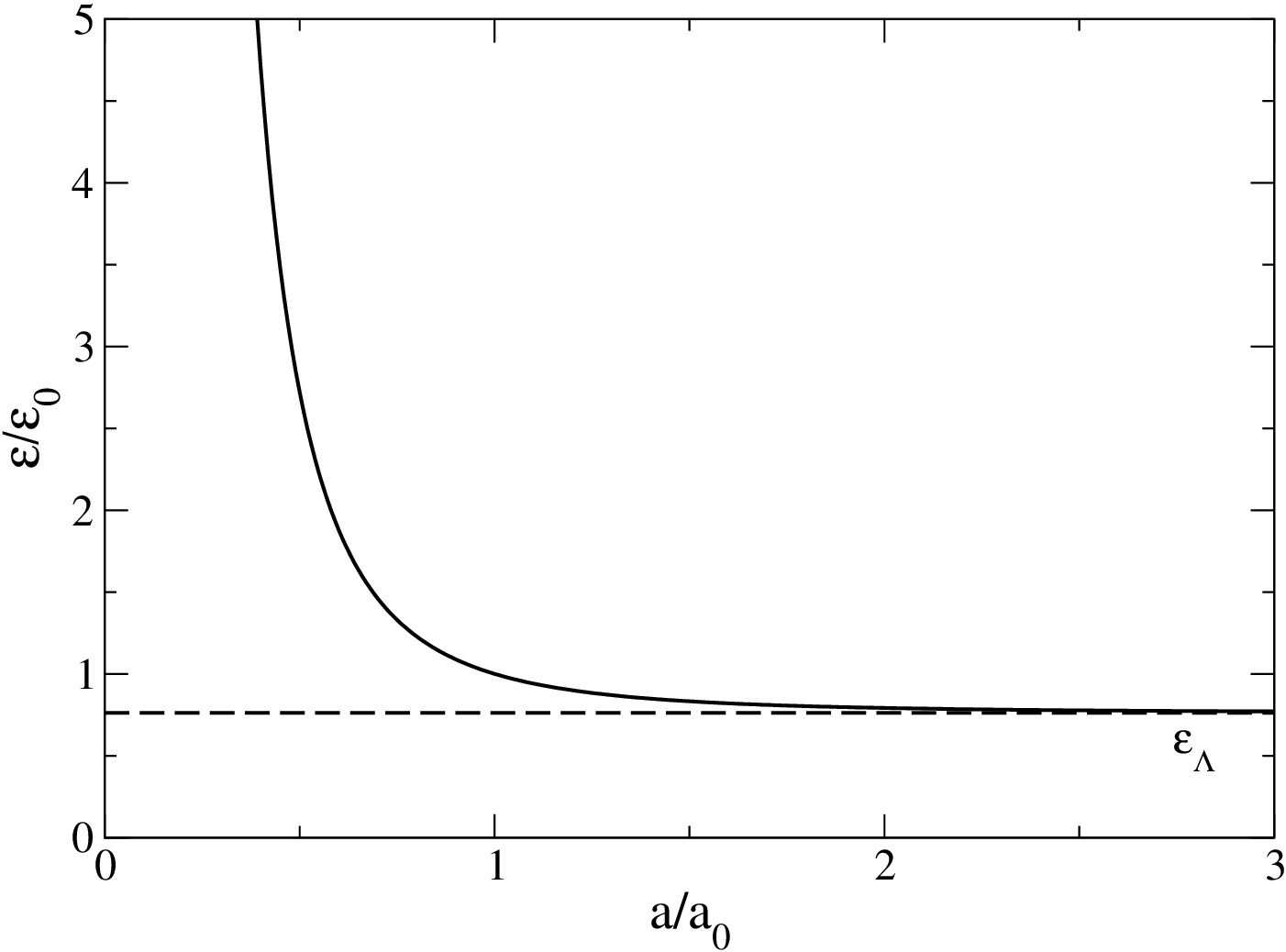}
\caption{Energy density as a function of the scale factor. We have taken
$\Omega_{m,0}=0.237$,
$\Omega_{\Lambda,0}=0.763$, and $\Omega_{s,0}=10^{-3}$ (we have
chosen a relatively large value of the density of stiff matter
$\Omega_{s,0}=10^{-3}$ for a better illustration of the results).\label{aepsPP}}
\end{figure}

\begin{figure}[!ht]
\includegraphics[width=0.98\linewidth]{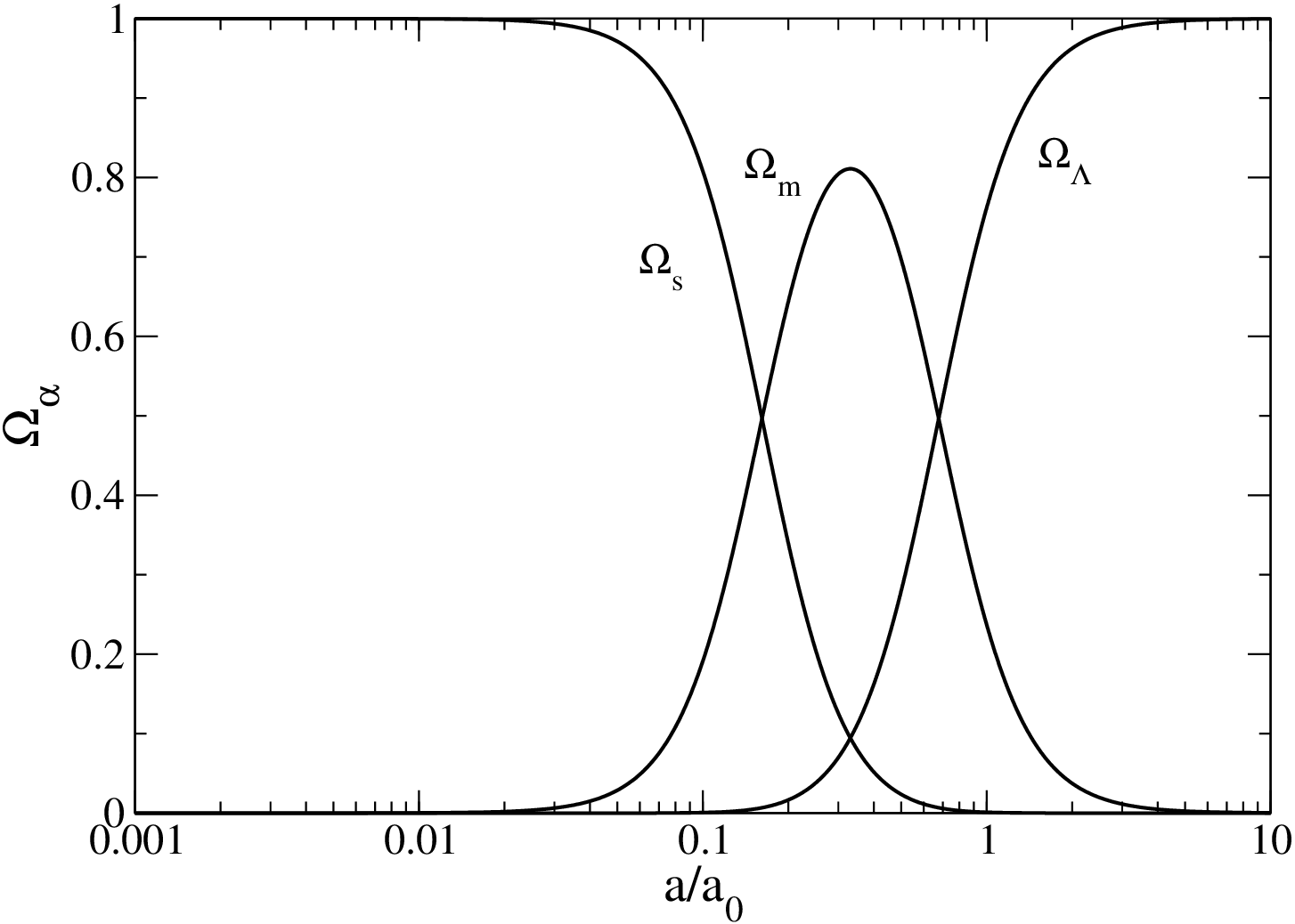}
\caption{Evolution of the proportion
$\Omega_{\alpha}=\epsilon_{\alpha}/\epsilon$ of the different components of the
universe with the scale factor. \label{proportionsPP}}
\end{figure}

We consider a universe made of stiff matter, radiation, dust matter and dark
energy treated as non-interacting species. Summing the contribution of each
species, the total energy density can by written as
\begin{equation}
\epsilon=\frac{\epsilon_{s,0}}{(a/a_0)^6}+\frac{\epsilon_{rad,0}}{(a/a_0)^4}+\frac{\epsilon_{m,0}}{(a/a_0)^3}+\epsilon_{\Lambda}.
\label{stiff1}
\end{equation}
In this model, the stiff matter dominates in the early universe. This is
followed 
by the radiation era, by the dust matter era and, finally, by the dark energy
era. Writing $\epsilon_{\alpha,0}=\Omega_{\alpha,0}\epsilon_0$ for each
species, we get
\begin{equation}
\frac{\epsilon}{\epsilon_0}=\frac{\Omega_{s,0}}{(a/a_0)^6}+\frac{\Omega_{rad,0}}
{(a/a_0)^4}+\frac{\Omega_{m,0}}
{(a/a_0)^3}+\Omega_{\Lambda,0}.
\label{stiff1b}
\end{equation}
The energy density starts from $\epsilon=+\infty$ at $a=0$, decreases, and
tends to $\epsilon_{\Lambda}$ for $a\rightarrow
+\infty$. The relation between the energy density and the scale factor is 
shown in Fig. \ref{aepsPP}. The proportions of stiff matter, dust matter and
dark energy as a function of the scale factor are shown in Fig.
\ref{proportionsPP}. Using Eq. (\ref{stiff1b}), the Friedmann
equation (\ref{f3}) takes the form
\begin{equation}
\frac{H}{H_0}=\sqrt{\frac{\Omega_{s,0}}{(a/a_0)^6}+\frac{\Omega_{rad,0}}{(a/a_0)^4}+\frac{\Omega_{m,0}}{(a/a_0)^3}+\Omega_{\Lambda,0}}
\label{stiff2}
\end{equation}
with $\Omega_{s,0}+\Omega_{rad,0}+\Omega_{m,0}+\Omega_{\Lambda,0}=1$ and
$H_0=(8\pi G\epsilon_0/3c^2)^{1/2}$. Note that we have taken $\Lambda=0$ in Eq.
(\ref{f3}) and accounted for the effect of the cosmological constant in the dark
energy density $\epsilon_{\Lambda}$. We also note the relation
\begin{equation}
\frac{\epsilon}{\epsilon_0}=\left
(\frac{H}{H_0}\right )^2
\label{stiff3}
\end{equation}
that will be used later. The evolution of the scale factor is given by
\begin{equation}
\int_0^{a/a_0} \frac{dx}{x \sqrt{\frac{\Omega_{s,0}}{x^6}+\frac{\Omega_{rad,0}}{x^4}+\frac{\Omega_{m,0}}{x^3}+\Omega_{\Lambda,0}}}=H_0 t.
\label{stiff4}
\end{equation}

We first ignore radiation ($\Omega_{rad,0}=0$) and consider a universe made of
stiff matter, dust matter, and dark energy. In that case, the Friedmann equation
(\ref{stiff4}) reduces to
\begin{equation}
\int_0^{a/a_0} \frac{dx}{x \sqrt{\frac{\Omega_{s,0}}{x^6}+\frac{\Omega_{m,0}}{x^3}+\Omega_{\Lambda,0}}}=H_0 t.
\label{stiff5}
\end{equation}

Using the identity
\begin{eqnarray}
\int \frac{dx}{x\sqrt{\frac{a}{x^3}+\frac{b}{x^6}+c}}\qquad\qquad\qquad\qquad\nonumber\\
=\frac{1}{3\sqrt{c}}\ln\left\lbrack a+2cx^3+2\sqrt{c}\sqrt{b+ax^3+cx^6}\right\rbrack,
\label{stiff6}
\end{eqnarray}
Eq. (\ref{stiff5}) can be solved analytically to give
\begin{eqnarray}
\frac{a}{a_0}=\Biggl \lbrack \left (\frac{\Omega_{m,0}}{\Omega_{\Lambda,0}}+2\sqrt{\frac{\Omega_{s,0}}{\Omega_{\Lambda,0}}}\right )\sinh^2\left (\frac{3}{2}\sqrt{\Omega_{\Lambda,0}}H_0 t\right )\nonumber\\
+\sqrt{\frac{\Omega_{s,0}}{\Omega_{\Lambda,0}}}\left (1-e^{-3\sqrt{\Omega_{\Lambda,0}}H_0 t}\right )\Biggr\rbrack^{1/3}.
\label{stiff7}
\end{eqnarray}
From Eq. (\ref{stiff7}), we can compute $H=\dot a/a$ leading to
\begin{eqnarray}
\left (\frac{a}{a_0}\right )^3\frac{H}{H_0}&=&\left
(\frac{\Omega_{m,0}}{2\sqrt{\Omega_{\Lambda,0}}}+\sqrt{\Omega_{s,0}}\right
)\sinh\left (3\sqrt{\Omega_{\Lambda,0}} H_0 t\right )\nonumber\\
&+&\sqrt{\Omega_{s,0}}
e^{-3\sqrt{\Omega_{\Lambda,0}}H_0 t}.
\label{stiff9}
\end{eqnarray}
The energy density is given by Eq. (\ref{stiff3})
where $H/H_0$ can be obtained from  (\ref{stiff9}) with Eq. (\ref{stiff7}). 

\begin{figure}[!ht]
\includegraphics[width=0.98\linewidth]{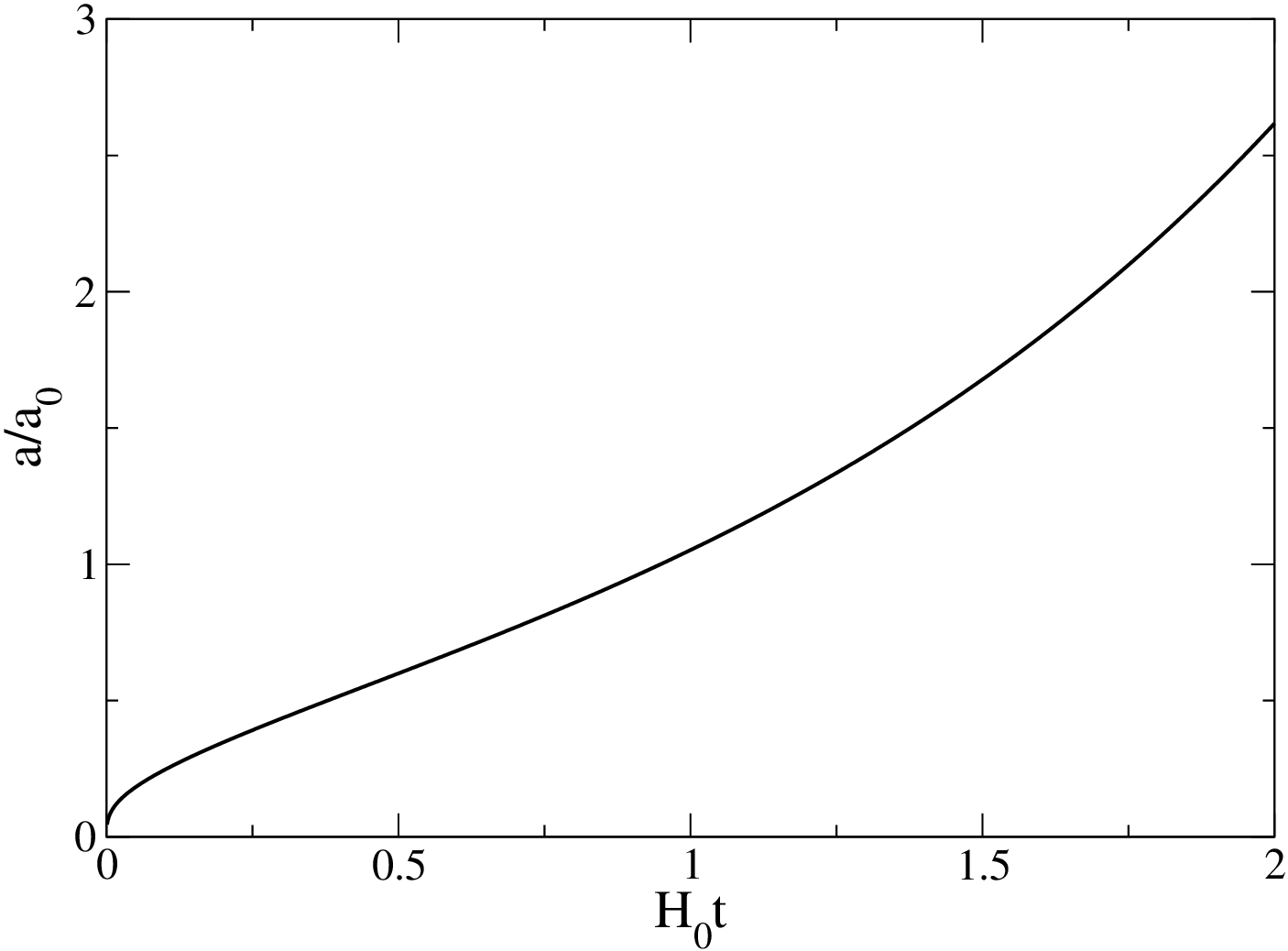}
\caption{Evolution of the scale factor as a function of time. \label{taPP}}
\end{figure}

\begin{figure}[!ht]
\includegraphics[width=0.98\linewidth]{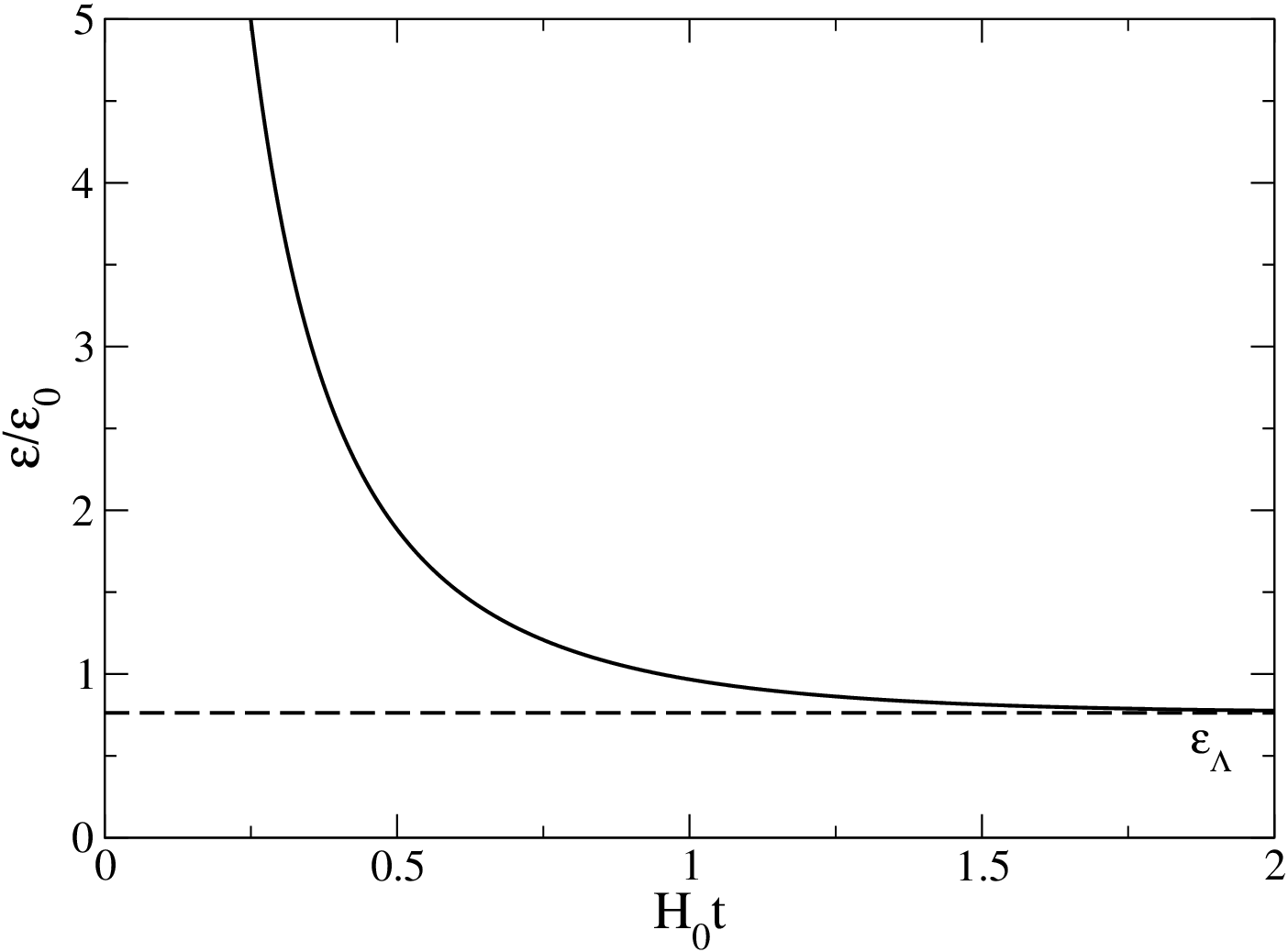}
\caption{Evolution of the energy density as a function of time. \label{tepsPP}}
\end{figure}

At $t=0$ the universe starts from a singular state at
which the scale factor $a=0$
while the energy density $\epsilon=+\infty$.  The scale factor increases with
time. For $t\rightarrow
+\infty$, we obtain
\begin{eqnarray}
\frac{a}{a_0}\sim \left (\frac{\Omega_{m,0}}{\Omega_{\Lambda,0}}+2\sqrt{\frac{\Omega_{s,0}}{\Omega_{\Lambda,0}}}\right )^{1/3}\frac{1}{2^{2/3}} e^{\sqrt{\Omega_{\Lambda,0}}H_0 t}.
\label{stiff8}
\end{eqnarray}
The energy density decreases with time and tends to
$\epsilon_{\Lambda}$ for $t\rightarrow +\infty$.
The expansion is decelerating during the stiff matter era and the dust matter
era while it is accelerating during the dark energy era. The
temporal evolution of the scale
factor and of the energy density are shown in  
Fig. \ref{taPP} and \ref{tepsPP}.

We consider a universe made of stiff matter and dust matter. In the absence of
dark energy ($\Omega_{\Lambda,0}=0$), using the identity
\begin{equation}
\int \frac{dx}{x\sqrt{\frac{a}{x^3}+\frac{b}{x^6}}}=\frac{2}{3a}\sqrt{b+ax^3},
\label{stiff10}
\end{equation}
we obtain 
\begin{eqnarray}
\frac{a}{a_0}=\left (\frac{9}{4}\Omega_{m,0}H_0^2 t^2+3\sqrt{\Omega_{s,0}}H_0 t\right )^{1/3},
\label{stiff11}
\end{eqnarray}
\begin{eqnarray}
\frac{\epsilon}{\epsilon_0}=\frac{4}{9H_0^2t^2}\left (\frac{1+\frac{2\sqrt{\Omega_{s,0}}}{3\Omega_{m,0}H_0 t}}{1+\frac{4\sqrt{\Omega_{s,0}}}{3\Omega_{m,0}H_0 t}}\right )^2.
\label{stiff12}
\end{eqnarray}

We consider a universe made of stiff matter and dark energy. In the absence of
matter ($\Omega_{m,0}=0$), using the identity
\begin{equation}
\int \frac{dx}{x\sqrt{\frac{b}{x^6}+c}}=\frac{1}{3\sqrt{c}}\ln\left\lbrack 2cx^3+2\sqrt{c}\sqrt{b+cx^6}\right\rbrack,
\label{stiff13}
\end{equation}
or setting $X=b/cx^6$ and using the identity
\begin{equation}
\int \frac{dX}{X\sqrt{X+1}}=\ln\left (\frac{\sqrt{1+X}-1}{\sqrt{1+X}+1}\right ),
\label{stiff14}
\end{equation}
we get
\begin{eqnarray}
\frac{a}{a_0}=\left (\frac{\Omega_{s,0}}{\Omega_{\Lambda,0}}\right )^{1/6}\sinh^{1/3}\left (3 \sqrt{\Omega_{\Lambda,0}}H_0 t\right ),
\label{stiff15}
\end{eqnarray}
\begin{eqnarray}
\frac{\epsilon}{\epsilon_0}=\frac{\Omega_{\Lambda,0}}{\tanh^2\left (3\sqrt{\Omega_{\Lambda,0}}H_0 t\right )}.
\label{stiff16}
\end{eqnarray}

We consider a universe made of stiff matter. In the absence of
dust matter and dark energy  ($\Omega_{m,0}=\Omega_{\Lambda,0}=0$), we find that
\begin{eqnarray}
\frac{a}{a_0}=\left (3\sqrt{\Omega_{s,0}}H_0 t\right )^{1/3},\qquad \frac{\epsilon}{\epsilon_0}=\frac{1}{9H_0^2t^2}.
\label{stiff17}
\end{eqnarray}

We consider a universe made of dust matter and dark energy. In the absence of
stiff matter ($\Omega_{s,0}=0$), using the identity
\begin{equation}
\int \frac{dx}{x\sqrt{\frac{a}{x^3}+c}}=\frac{1}{3\sqrt{c}}\ln\left\lbrack a+2cx^3+2\sqrt{c}\sqrt{ax^3+cx^6}\right\rbrack,
\label{stiff18}
\end{equation}
or setting $X=a/cx^3$ and using Eq. (\ref{stiff14}), we obtain
\begin{eqnarray}
\frac{a}{a_0}=\left (\frac{\Omega_{m,0}}{\Omega_{\Lambda,0}}\right )^{1/3}\sinh^{2/3}\left (\frac{3}{2}\sqrt{\Omega_{\Lambda,0}}H_0 t\right ),
\label{stiff19}
\end{eqnarray}
\begin{eqnarray}
\frac{\epsilon}{\epsilon_0}=\frac{\Omega_{\Lambda,0}}{\tanh^2\left (\frac{3}{2}\sqrt{\Omega_{\Lambda,0}}H_0 t\right )}.
\label{stiff20}
\end{eqnarray}
This solution  coincides with the $\Lambda$CDM model.

We consider a universe made of dark energy. In the absence of stiff matter and
dust matter ($\Omega_{s,0}=\Omega_{m,0}=0$), we obtain 
\begin{eqnarray}
a(t)=a(0)e^{\sqrt{\frac{\Lambda}{3}}t},\qquad
\epsilon=\epsilon_{\Lambda}.
\label{stiff21}
\end{eqnarray}
This is de Sitter's solution. 

We consider a universe made of dust matter. In the absence of stiff matter and
dark energy  ($\Omega_{s,0}=\Omega_{\Lambda,0}=0$), we obtain 
\begin{eqnarray}
\frac{a}{a_0}=\left (\frac{9}{4}\Omega_{m,0}H_0^2 t^2\right )^{1/3},\qquad \frac{\epsilon}{\epsilon_0}=\frac{4}{9H_0^2t^2}.
\label{stiff22}
\end{eqnarray}
This is the Einstein-de Sitter (EdS) solution.

We now come back to the general equation (\ref{stiff2}) including the
contribution of radiation.

The transition between the stiff matter era and the radiation era is obtained by
taking $\Omega_{m,0}=\Omega_{\Lambda,0}=0$ in Eq. (\ref{stiff2}). In that case,
the integral in Eq. (\ref{stiff4}) can be performed analytically leading to
\begin{eqnarray}
2\sqrt{\Omega_{rad,0}}\frac{a}{a_0}\sqrt{\Omega_{s,0}+\Omega_{rad,0}\left (\frac{a}{a_0}\right )^2}-2\Omega_{s,0}\nonumber\\
\times\ln\left\lbrack \Omega_{rad,0}\frac{a}{a_0}+\sqrt{\Omega_{rad,0}}\sqrt{\Omega_{s,0}+\Omega_{rad,0}\left (\frac{a}{a_0}\right )^2}\right\rbrack\nonumber\\
+\Omega_{s,0}\ln(\Omega_{s,0}\Omega_{rad,0})=4(\Omega_{rad,0})^{3/2} H_0 t.\nonumber\\
\label{stiff23}
\end{eqnarray}

The transition between the radiation era and the matter era is obtained by
taking $\Omega_{s,0}=\Omega_{\Lambda,0}=0$ in Eq. (\ref{stiff2}). In that case,
the integral in Eq. (\ref{stiff4})  can be performed analytically leading
to\footnote{We have determined the constant of integration in Eq.
(\ref{stiff24}) such that $a=0$ at $t=0$. This implicitly assumes that there is
no stiff matter in the early universe. Otherwise, we need to determine the
constant of integration by matching the solution (\ref{stiff24}) with the
solution (\ref{stiff17}) of the stiff matter era.}
\begin{eqnarray}
H_0 t=-\frac{2}{3}\frac{1}{(\Omega_{m,0})^{1/2}}\left (\frac{2\Omega_{rad,0}}{\Omega_{m,0}}-\frac{a}{a_0}\right )\sqrt{\frac{\Omega_{rad,0}}{\Omega_{m,0}}+\frac{a}{a_0}}\nonumber\\
+\frac{4}{3}\frac{(\Omega_{rad,0})^{3/2}}{(\Omega_{m,0})^2}.\qquad
\label{stiff24}
\end{eqnarray}
 Eq. (\ref{stiff24}) can also be written as
\begin{equation}
\left (\frac{a}{a_0}\right )^3-3\frac{\Omega_{rad,0}}{\Omega_{m,0}}\left (\frac{a}{a_0}\right )^2=\frac{9}{4}\Omega_{m,0}H_0^2t^2-6\frac{\Omega_{rad,0}^{3/2}}{\Omega_{m,0}}H_0 t.
\label{stiff25}
\end{equation}
This is a cubic equation for $a/a_0$.  

For mathematical completeness, we also give the equations corresponding to a
universe containing only radiation and dark energy
($\Omega_{s,0}=\Omega_{m,0}=0$). 
They write
\begin{eqnarray}
\frac{a}{a_0}=\left (\frac{\Omega_{rad,0}}{\Omega_{\Lambda,0}}\right )^{1/4}\sinh^{1/2}\left (2\sqrt{\Omega_{\Lambda,0}}H_0 t\right ),
\label{stiff26}
\end{eqnarray}
\begin{eqnarray}
\frac{\epsilon}{\epsilon_0}=\frac{\Omega_{\Lambda,0}}{\tanh^2\left (2\sqrt{\Omega_{\Lambda,0}}H_0 t\right )}.
\label{stiff27}
\end{eqnarray}
For a  universe containing only radiation ($\Omega_{s,0}=\Omega_{m,0}=\Omega_{\Lambda,0}=0$) we get
\begin{eqnarray}
\frac{a}{a_0}=\Omega_{rad,0}^{1/4}\sqrt{2H_0 t}, \qquad \frac{\epsilon}{\epsilon_0}=\frac{1}{(2H_0 t)^2}.
\label{stiff28}
\end{eqnarray}

Finally, we can propose a simple generalization of Eq. (\ref{stiff2}) that
includes a phase of early inflation. Using the arguments developed in
\cite{cosmopoly1,cosmopoly2} we get
\begin{equation}
\frac{H}{H_0}=\sqrt{\frac{\Omega_{s,0}}{(a/a_0)^6+(a_1/a_0)^6}+\frac{\Omega_{rad,0}}{(a/a_0)^4}+\frac{\Omega_{m,0}}{(a/a_0)^3}+\Omega_{\Lambda,0}},
\label{stiff29}
\end{equation}
where the constant $a_1$ is determined by the relation $\epsilon_P
a_1^6=\epsilon_{s,0} a_0^6$ where $\epsilon_P=\rho_P c^2$ is the Planck energy
density. The transition between the inflation era and the stiff matter era is
obtained by taking $\Omega_{rad,0}=\Omega_{m,0}=\Omega_{\Lambda,0}=0$ in Eq.
(\ref{stiff29}). In that case, Eq. (\ref{stiff29})  can be
integrated analytically to give \cite{cosmopoly1}:
\begin{equation}
\sqrt{R^6+1}-\ln\left (\frac{1+\sqrt{R^6+1}}{R^3}\right )=3Kt+C,
\label{stiff30}
\end{equation}
where we have defined  $R=a/a_1$ and $K=H_0(\Omega_{s,0} a_0^6/a_1^6)^{1/2}=(8\pi G\rho_P/3)^{1/2}$. $C$ is a constant of integration that can be determined by requiring that $a=l_P$ at $t=0$ where $l_P$ is the Planck length.

\end{document}